\documentclass[manuscript]{aastex}

\shorttitle{OPTICAL PROPERTIES OF THE ULX HoIX~X-1}
\shortauthors{Gris\'e et al.}

\usepackage{subfig}
\usepackage{natbib}
\usepackage{booktabs}
\usepackage{color}
\usepackage{graphicx}
\usepackage{threeparttable}

\begin{document}


\title{OPTICAL PROPERTIES OF THE ULTRALUMINOUS X-RAY SOURCE HOLMBERG IX X-1 AND ITS STELLAR ENVIRONMENT}


\author{F. Gris\'e\altaffilmark{1}, P. Kaaret\altaffilmark{1}, M. W. Pakull\altaffilmark{2}, and C. Motch\altaffilmark{2}}
\altaffiltext{1}{Department of Physics and Astronomy, University of Iowa,
    Van Allen Hall,\\Iowa City, IA 52242, USA}

\email{fabien-grise@uiowa.edu}
\altaffiltext{2}{Observatoire Astronomique de Strasbourg, 11 rue de l'Universit\'e, 67000 Strasbourg, France}






\begin{abstract}
Holmberg~IX~X-1 is an archetypal ultraluminous X-ray source (ULX). Here we study the properties of the optical counterpart and of its stellar environment using optical data from SUBARU/Faint Object Camera and Spectrograph, {\it GEMINI/GMOS-N} and {\it Hubble Space Telescope} (HST)/Advanced Camera for Surveys, as well as simultaneous {\it Chandra} X-ray data. The $V \sim 22.6$ spectroscopically identified optical counterpart is part of a loose cluster with an age $\la 20\ \mathrm{Myr}$. Consequently, the mass upper limit on individual stars in the association is about $20\ M_{\mathrm{\odot}}$. The counterpart is more luminous than the other stars of the association, suggesting a non-negligible optical contribution from the accretion disk. An observed UV excess also points to non-stellar light similar to X-ray active low-mass X-ray binaries. A broad \ion{He}{2}$\lambda$4686 emission line identified in the optical spectrum of the ULX further suggests optical light from X-ray reprocessing in the accretion disk. Using stellar evolutionary tracks, we have constrained the mass of the counterpart to be $\ga 10\ M_{\odot}$, even if the accretion disk contributes significantly to the optical luminosity. Comparison of the photometric properties of the counterpart with binary models show that the donor may be more massive, $\ga 25 M_{\mathrm{\odot}}$, with the ULX system likely undergoing case AB mass transfer. Finally, the counterpart exhibits photometric variability of 0.14 mag between two {\it HST} observations separated by 50 days which could be due to ellipsoidal variations and/or disk reprocessing of variable X-ray emission.
\end{abstract}



\section{Introduction}

Ultraluminous X-ray sources (ULXs) are compact X-ray sources that are located away from the nucleus of their host galaxy, emitting well above the Eddington limit of a $20\ M_{\mathrm{\sun}}$ black~hole ($L_\mathrm{X}\sim 3\times 10^{39}\ \mathrm{erg\ s^{-1}}$) assuming isotropic emission. It is now agreed that ULXs are accreting systems powered by black holes, due to the high X-ray variability they sometimes display on timescales as short as minutes. Among the remaining questions is the nature of these black holes: are they stellar-mass black holes similar to Galactic X-ray binaries but with higher accretion rates or do they contain intermediate-mass black holes? See \citet{2007ApSS.311..203R} for a recent X-ray review.

Holmberg~IX~X-1 (HoIX~X-1 also known as M81~X-9) is a prototype ULX located in Holmberg~IX, an irregular dwarf galaxy cataloged by \citet{1974A&AS...18..463H} belonging to the M81 group of galaxies. It is located at about 3.6 Mpc (the distance of M81; \citealt{1994ApJ...427..628F}), corresponding to a distance modulus $m-M=27.78$. Holmberg~IX may be a tidal dwarf galaxy resulting from the major interaction between M81, M82, and NGC~3077 that occurred 200~Myr ago or an old dwarf galaxy \citep{2008ApJ...676L.113S}. Indeed, this group of galaxies is located in a giant \ion{H}{1} cloud with tidal bridges connecting the different components \citep{1994Natur.372..530Y}. Numerous optical knots have been detected along the \ion{H}{1} filaments where there are overdensities, Holmberg~IX being one of the most prominent such objects.

Discovered by Einstein \citep{1988ApJ...325..544F} and studied with all the following X-ray satellites (see \citet{2010MNRAS.403.1206V} for an X-ray review), HoIX~X-1 is an extremely luminous X-ray source with an isotropic X-ray luminosity (0.3--10 keV) usually above $10^{40}\ \mathrm {erg\ s^{-1}}$ and thus is one of the most luminous ULXs. The X-ray spectrum of this ULX has been interpreted in two different ways, either as a cool accretion disk ($kT_{\mathrm{in}} \sim 0.25\ \mathrm{keV}$) with an additional power law or as a slim disk characterized by a hot thermal component ($kT_{\mathrm{in}} \sim \mathrm{1.4-1.8\ keV}$). This has led to two opposite conclusions with the former interpretation suggesting the presence of an intermediate-mass black hole ($M \sim 1000\ M_{\mathrm{\odot}}$) accreting at a sub-Eddington rate \citep{2004ApJ...607..931M} and the latter implying a stellar-mass black hole ($M \sim 10\ M_{\mathrm{\odot}}$) accreting at a super-Eddington rate \citep{2006PASJ...58.1081T}. Recently, a model consisting of a disk plus a cool, optically thick corona \citep{2009MNRAS.397.1836G,2010MNRAS.403.1206V} has been applied to high-quality X-ray spectra of ULXs and in particular to HoIX~X-1.  This model suggests a stellar-mass black hole with super-Eddington accretion.

HoIX~X-1 is located about $2 \arcmin$ ($\sim 2.1\ \mathrm{kpc}$) northeast of the center of Holmberg~IX and about $1 \arcmin$ ($\sim 1\ \mathrm{kpc}$) outside of what we can consider the limit of the stellar envelope of the galaxy. This ULX is thus present in a zone where there is no intense stellar formation, but we note the existence of a close molecular cloud \citep{1992A&A...262L...5B} at $40 \arcsec$ southeast of its position. The Galactic interstellar extinction is low, E($B-V$) $\sim$ 0.08 \citep{1998ApJ...500..525S}, making this source a very good target for an optical photometric and spectroscopic study.

The striking characteristic of HoIX~X-1 at optical wavelengths is the presence of a huge ionized nebula, see \citet{1995ApJ...446L..75M} and Figure~\ref{hoixx1_images_neb_couleur}, with a size of $300 \times 470$ pc at the distance of 3.6 Mpc. This and further work \citep{2002astro.ph..2488P,2003RMxAC..15..197P} has shown that shocks should play a key role in the nebula even if photoionization from the X-ray source is possibly required to understand all the properties of the nebula \citep{2008RMxAA..44..301A}. The optical counterpart of the ULX is the most luminous object belonging to a small stellar association located inside the nebula \citep{2006IAUS..230..302G,2006ApJ...641..241R}. This paper concentrates on the properties of the optical counterpart and its stellar environment, revising and expanding our previous results \citep{2006IAUS..230..302G}.

\section{Observations and data analysis}

\subsection{SUBARU}

The SUBARU data set comes from two observing programs (PIs: T. Tsuru, M. Pakull) and makes use of the SUBARU/Faint Object Camera and Spectrograph (FOCAS) instrument \citep{2002PASJ...54..819K}. This instrument has been used to study the counterparts and environment of several ULXs, one of them being HoIX~X-1. This data set consists of imaging through broadband filters as {\it B}, {\it V}, and {\it R} and through narrowband filters as H$_\alpha$ and [\ion{O}{3}], as well as spectroscopy. All the raw frames were reduced under {\small ESO MIDAS} using standard procedures, i.e., correcting for bias and flat fields. The two spectra were combined and then wavelength and flux calibrated.

The local environment to the ULX is not completely resolved, so point-spread function (PSF) fitting photometry is necessary to separate blended objects. We used the package {\small DAOPHOT II} \citep{1990ASPC....8..289S} under {\small ESO MIDAS} and we employed a quadratically varying PSF in the field.

Our absolute photometry in the {\it B}, {\it V}, and {\it R} bands uses three consecutive observations made on 2003 Jan 26 (see Table \ref{tab_subarudata} for details of the SUBARU observations). Photometric transformations were derived with the only observed standard field, Ru152, for which we used both the Landolt standard stars and also a set of stars defined by Peter Stetson \footnote{Photometry of this secondary set of stars is available on the CADC Web site, http://www3.cadc-ccda.hia-iha.nrc-cnrc.gc.ca/community/STETSON/standards/}. This permitted us to obtain a more homogeneous calibration on the range of photometric colors and also to use stars fainter than the 14th mag.

To check our photometry, we used the Sloan Digital Sky Survey (SDSS) data release 6 \citep{2008ApJS..175..297A} which imaged fields around our target. Thanks to its 2.5 m telescope, photometry with an accuracy of 2\%--3\% is possible down to 21--22nd magnitude.
The filters' system used by the SDSS is noticeably different from the Johnson-Cousins system, so we had to transform the $u$, $g$ and $r$ magnitudes into $B$, $V$, $R$. For that we used the empirical transformations of Lupton\footnote{See http://sdss.org/dr6/algorithms/sdssUBVRITransform.html} (2005) who made a correlation between SDSS photometry and Stetson photometry. A possible dependence on color or metallicity related to the spectral type of stars has not been taken into account. As our goal is to make a global comparison between the two photometric data sets and not an analysis of SDSS data, we can content ourselves with this approach.

\begin{figure*}
   \begin{tabular}{cc}
      \includegraphics[width=8cm]{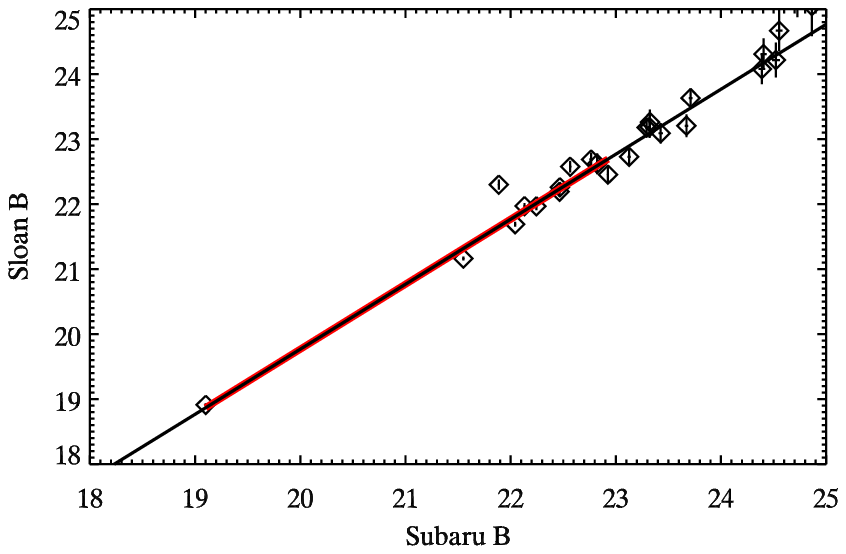}
     &\includegraphics[width=8cm]{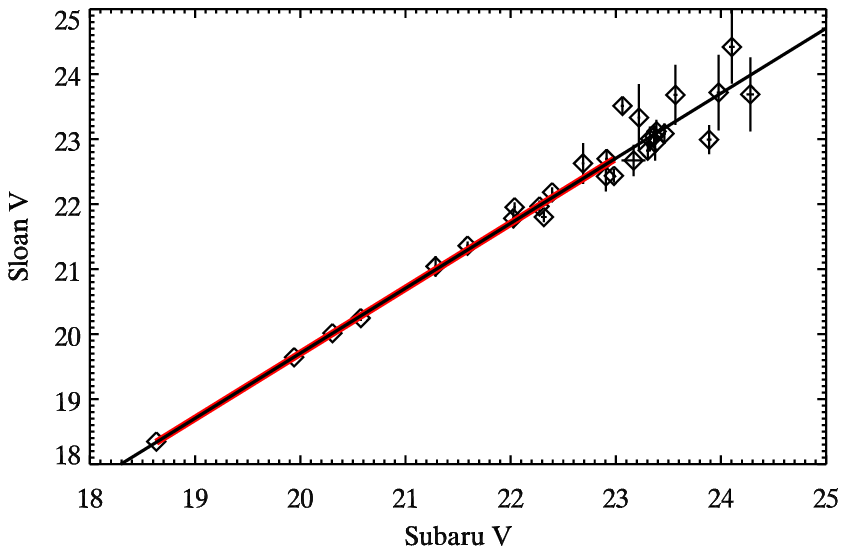}\\
      \includegraphics[width=8cm]{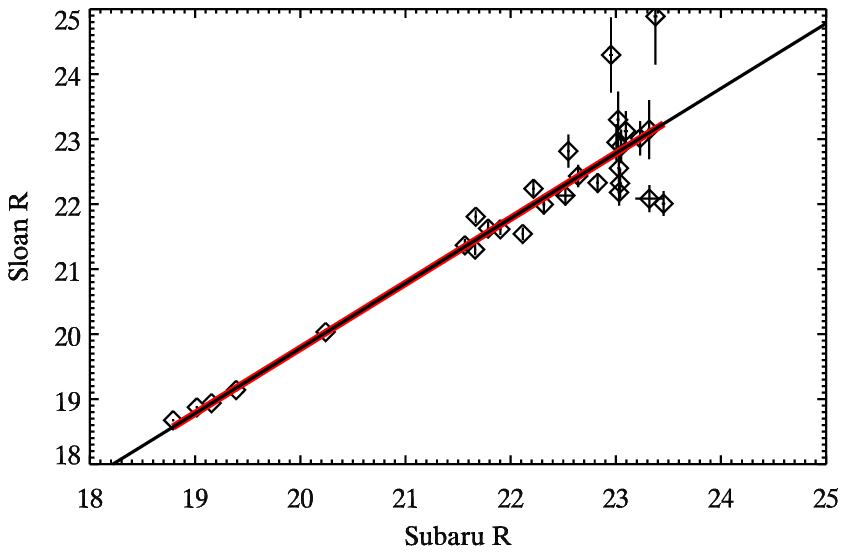}
     &\\
   \end{tabular}
   \caption[]{Photometric correlations between SUBARU data and the SDSS survey for three broadband filters ($B$, $V$, and $R$). The red line shows the range used to compute the photometric offset, using a weighted linear regression.}
   \label{comp_subaru_sloan}
\end{figure*}

The results of this comparison are shown in Figure \ref{comp_subaru_sloan}. We note the good linear agreement in the three filters. The scatter at faint magnitudes is consistent with the photometric errors of the SDSS and is more evident in the $R$ filter, where the errors are higher than in the other filters.
However, there is an offset in each filter, with the SUBARU magnitudes being fainter than the SDSS values (+0.23 mag in $B$, +0.30 mag in $V$, and +0.22 mag in $R$). The reason for these offsets are unclear. In the following, all SUBARU measurements have been corrected by these offsets.

Long-slit spectroscopy has been done with the slit centered on the assumed counterpart. The observational settings include a 150 lines $\mathrm{mm^{-1}}$ grism, an $SY47$ filter, and a slit of 0\farcs6 in width. This translates into an accessible wavelength range of 4600--7600 \AA, a dispersion of 2.8 \AA\ $\mathrm{pixel^{-1}}$ and a resolution $\lambda/\Delta\lambda = 500$ at 6500 \AA, i.e., the spectral resolution is approximately 8 \AA.

\subsection{{\it HST}}

The second data set used in this study comes from the {\it Hubble Space Telescope (HST)}/Advanced Camera for Surveys (ACS) archive (Program GO-9796 ; PI: J.\ Miller). Observations were made on 2004 February 07 with the Wide-Field Camera (WFC) in the $F435W$, $F555W$, and $F814W$ filters, and also with the High Resolution Camera (HRC) in the $F330W$ filter. One additional observation was made on 2004 March 25 with the WFC in the $F555W$ filter. More details about these observations are given in Table \ref{tab_hstdata}. We have re-analyzed the data using the latest calibration files available at the time of the analysis ({\small STSDAS} v3.5) and have applied the standard corrections, especially about the spatial distortion (Multidrizzle v2.7.2; \citealt{2002hstc.conf..337K}). As in the SUBARU observations, the field of interest looks moderately crowded in the {\it HST}/ACS image so we used again {\small DAOPHOT II} \citep{1990ASPC....8..289S} for the photometric analysis. This analysis is similar to our previous study of the ULX NGC~1313~X-2 and details can be found in \citet[G08 hereafter]{2008A&A...486..151G}.

\subsection{Comparison between {\it HST} and SUBARU photometry}

Optical variability is an important parameter in studies of X-ray binaries, since it can constrain the orbital period of the system or quantify a possible variability linked to the accretion disk. To study photometric variability using different telescopes, we have to make sure that the magnitudes derived from the {\it HST} and SUBARU data sets are consistent. As for our previous study (G08), we have isolated bright point-like sources in the two data sets and compared their luminosities to study the photometric accuracy of our two data sets.

\begin{figure*}
   \begin{tabular}{cc}
      \resizebox{8cm}{!}{\includegraphics{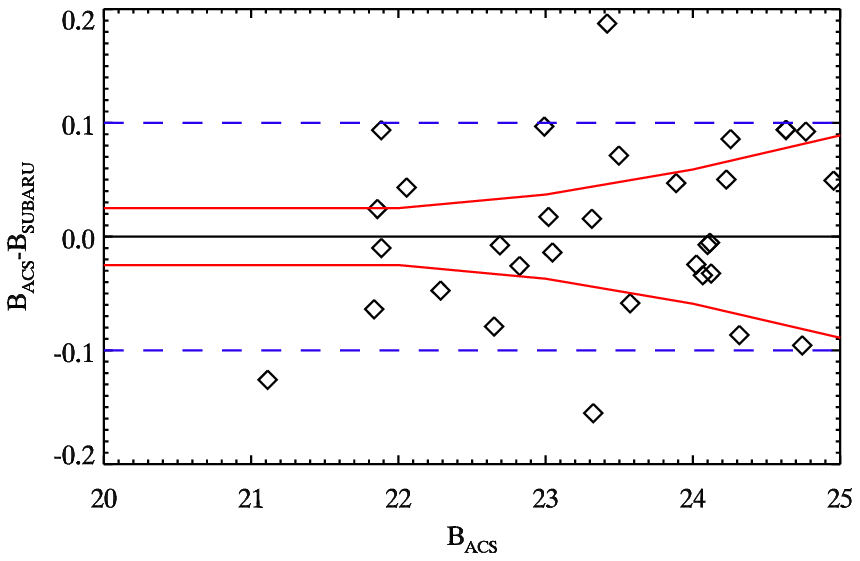}}
     &\resizebox{8cm}{!}{\includegraphics{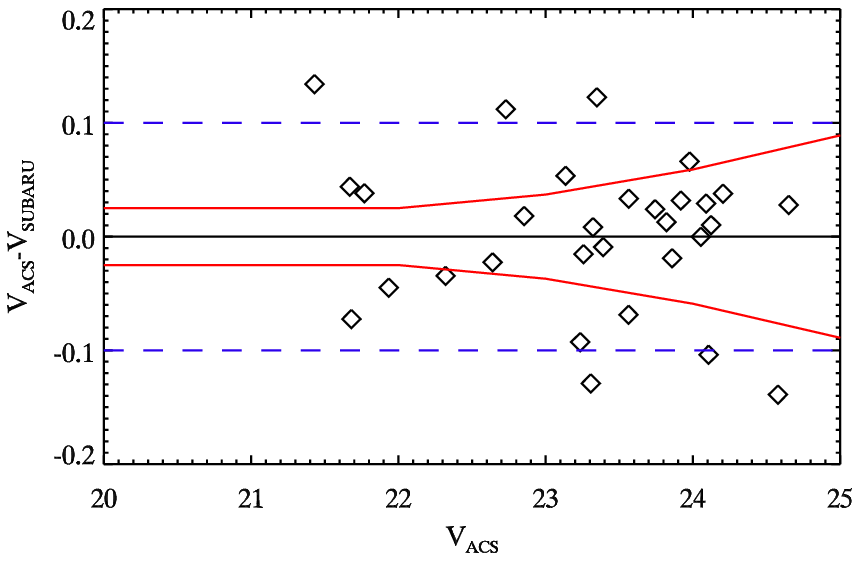}}\\
   \end{tabular}
   \caption{Difference between SUBARU/FOCAS and {\it HST}/ACS photometry 
   in the $B$ (left panel) and $V$ bands (right panel). For this comparison, 
   we only used bright, isolated stars. We find that 50\% of them lie 
   within the combined photometric errors of the {\it HST} and SUBARU observations 
   (solid red lines), and for most of the others the difference 
   is only $\le 0.1$ mag (dashed blue lines).}
   \label{comparison_hst_subaru}
\end{figure*}

Figure~\ref{comparison_hst_subaru} shows the magnitude differences between FOCAS and ACS in the $B$ and $V$ filters for observed stars. For 50\% of them, the differences are consistent with the combined photometric errors of the two instruments. For the other objects, the differences are almost always below 0.1 mag. Standard deviations are $\sigma_B = 0.06\ \mathrm{mag}$ and $\sigma_V = 0.07\ \mathrm{mag}$ which turn out to be the same as found in our previous study of NGC~1313~X-2 between {\it HST}/ACS and {\it ESO/VLT} data (G08). We note that deviations of the order of 0.1 mag are easily explained by, for example, a 23 mag object blended with a 26 mag object.

Photometric study of the ULX counterpart with SUBARU, Table~\ref{tab_magnitudes_subaru}, was done on 2003 Jan 26 with three consecutive observations in the three photometric bands $B$, $V$ and $R$, strongly limiting the risk of seeing intrinsic variability which could have introduced a bias in colors estimation. Comparing with {\it HST} data, Table~\ref{tab_magnitudes_hst_hoix}, we note excellent agreement in the ($B-V$) colors with magnitude differences being explained by variability.

\subsection{{\it GEMINI}}

The last optical data set used in this study comes from the {\it GEMINI/GMOS-N} archive (Program ID GN-2008A-Q-49; PI: J.\ Gladstone). The observations consist of six spectra taken consecutively in the night of 2008 March 07 with the {\it Gemini Multi-Object Spectrograph} ({\it GMOS}) \citep{2004PASP..116..425H} installed on the 8.1 m Gemini North telescope. The B600 grism was used, covering a wavelength range of 3160--6070 \AA\ with a dispersion of 0.90 \AA\ $\mathrm{pixel^{-1}}$ including a $2 \times 2$ binning, and the slit width was $0\farcs75$.

Data reduction was performed using the latest {\it GEMINI} IRAF package v1.10. All standard tools to reduce long-slit spectroscopy are included and easily handle the three CCDs frames. Flat fields were created and used with the processed bias to correct for instrumental defects. After wavelength calibration, the six spectra were co-added using the ccdclip rejection algorithm to remove cosmic rays and other bad pixels. Careful extraction of the spectrum was done due to the trace which appeared to vary by six pixels from one side of the frame to the other. We did not perform an absolute flux calibration.

\subsection{{\it Chandra} data}

Simultaneous X-ray and optical observations are quite rare for ULXs, so we took the opportunity that two X-ray pointings were done at the same time as the {\it HST} observations. We retrieved the data from the {\it Chandra} archive which consist of two 5 ks ACIS-S observations (PI: J.\ Miller, see details in Table \ref{tab_chandradata}). We used the level 2 event files but were faced with severe pile-up due to the combination of the high count rate of the source and the frame time used, 1.8~s. Standard extraction of the data led to highly distorted spectra, and use of the pile-up model \citep{2001ApJ...562..575D} in Xspec 12.6.0l \citep{1996ASPC..101...17A} was not able to solve this problem. Thus, we extracted only the counts coming from the PSF wings of the source, where pile-up should be negligible. To do so, we used the recipe devised by M. Tsujimoto\footnote{\textrm{http://www.astro.isas.jaxa.jp/$\mathrm{\sim}$tsujimot/arfcorr.html}} that provides a method to correct the Ancillary Response File when using events from the outskirts of the PSF. To maximize the signal-to-noise ratio without including a large fraction of piled-up events, we used a series of different extraction annuli, then fitted an absorbed power law in Xspec and plotted the power-law index versus the radius of the inner extraction annulus. The power-law index is unphysically hard ($\Gamma \sim 0.5$) at small inner radii, but increases at larger radii until reaching a plateau. This is where pile-up becomes negligible and corresponds to an inner extraction radius of 3.5 and 4.0 pixels for the first and second observation, respectively. The outer radius for the extraction was taken as 20 pixels for both observations.

Results of the spectral fits, involving 286 and 240 counts, respectively, are shown in Table~\ref{tab_chandradata}, where the spectra were grouped to have bins of at least 20 counts. We were unable to use models more complex than an absorbed power law due to the loss of counts resulting from the pile-up.
Our fits, using a column density fixed to $N_{\mathrm{H}} = 1.0 \times 10^{21}\ \mathrm{cm^{-2}}$ \citep{2009MNRAS.397.1836G,2010MNRAS.403.1206V} are quite similar for the two observations with a power-law index of $\sim 1.8$ and an absorbed flux of $\sim 1.1 \times 10^{-11}\ \mathrm{erg\ s^{-1}\ cm^{-2}}$. This is consistent with the apparent permanent hard spectrum of HoIX~X-1 \citep{2009ApJ...702.1679K}.

\section{Results and discussion}

\subsection{Identification of the optical counterpart}

\subsubsection{Photometric identification}

\begin{figure*}
   \begin{tabular}{cc}
	\rotatebox{90}{\resizebox{8.5cm}{!}{\includegraphics{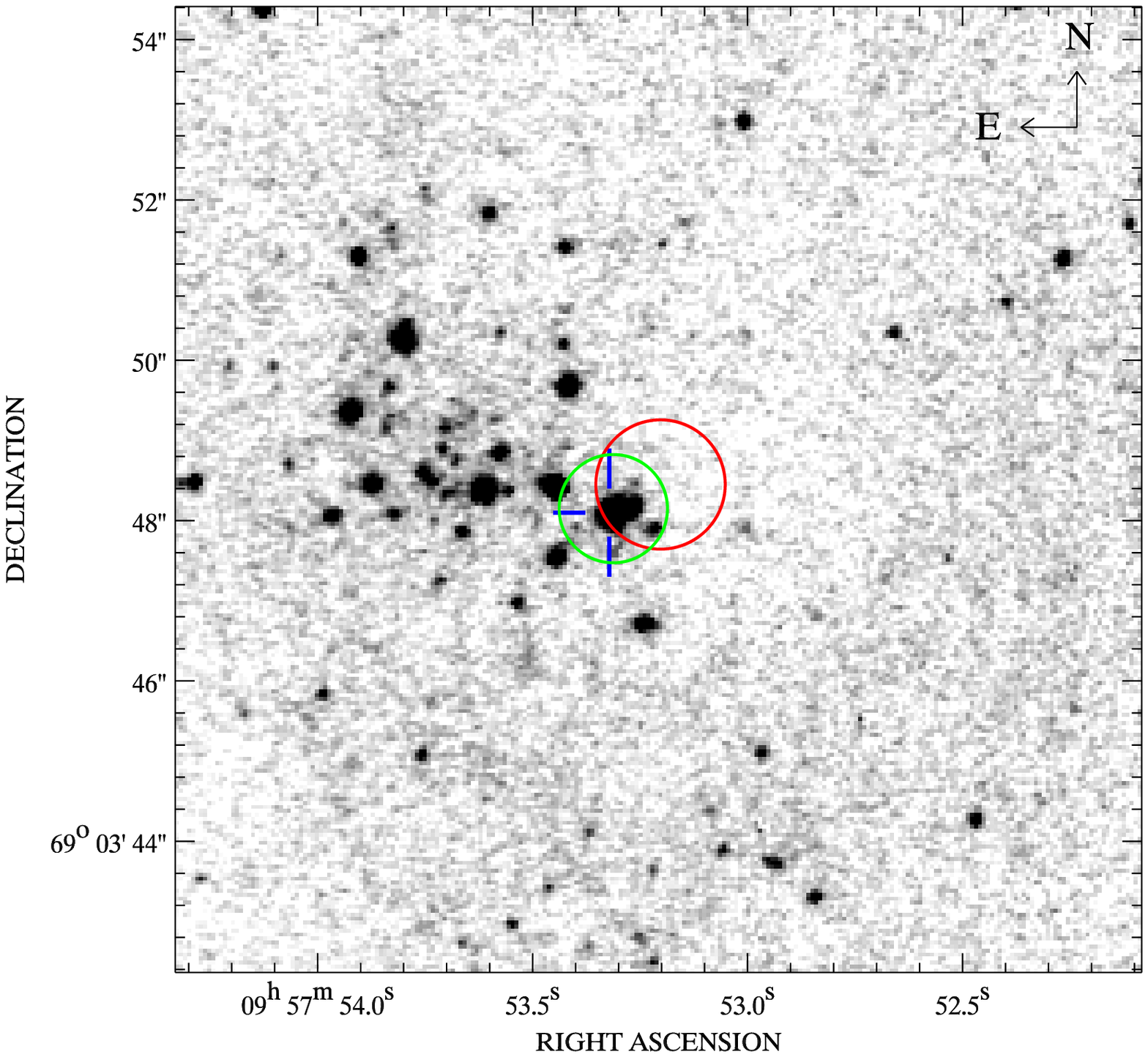}}}
     &	\rotatebox{90}{\resizebox{8.5cm}{!}{\includegraphics{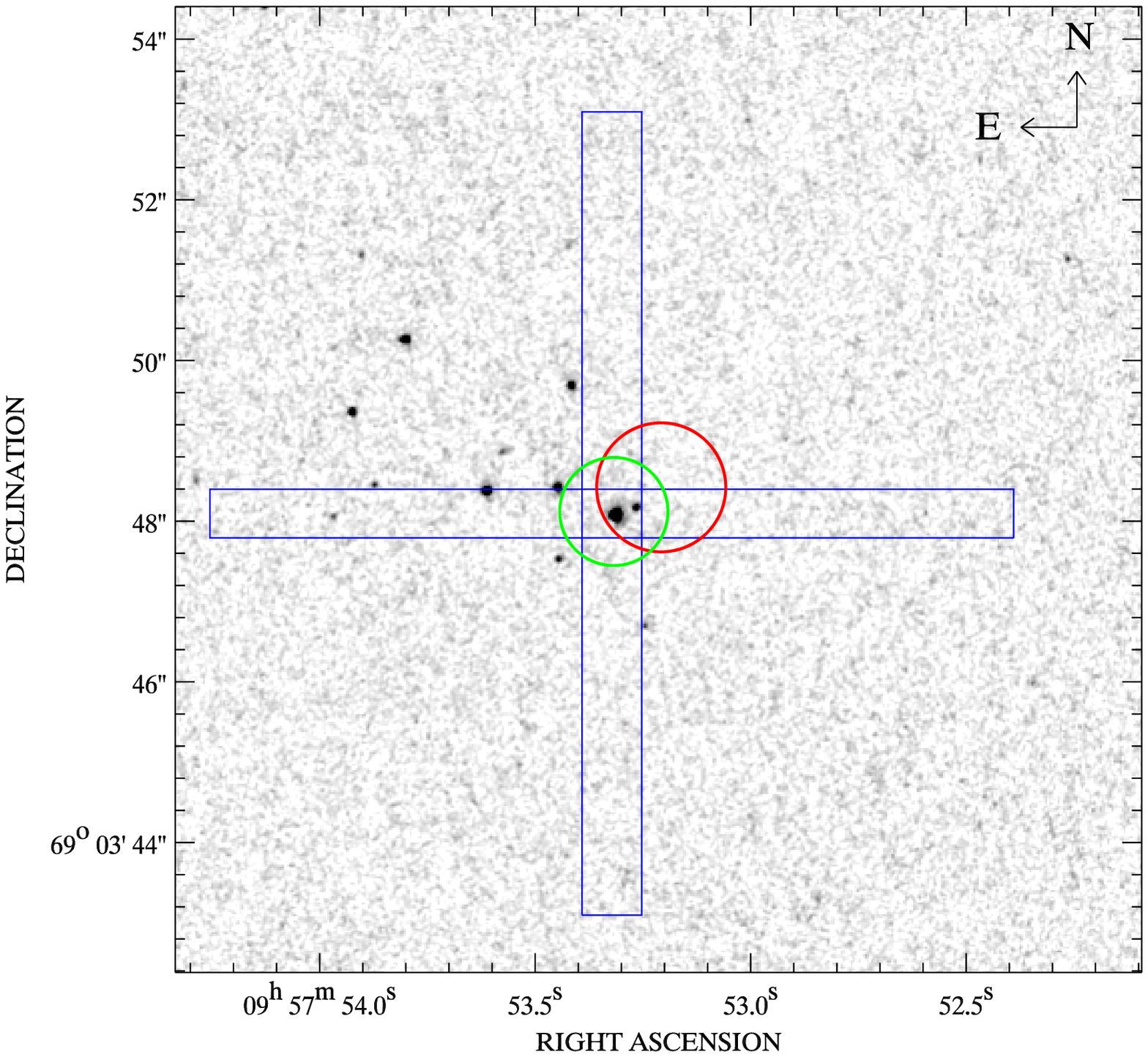}}}\\
   \end{tabular}
   \caption[]{Identification of HoIX~X-1 ULX on an {\it HST}/ACS image in the $F435W$ filter (left) and in the $F330W$ filter (right). The counterpart is designated by a blue cross (left). Slits from the SUBARU (position angle of 180\degr) and {\it GEMINI} observations (position angle of 90\degr) are overlaid on the right image.
The {\it Chandra} (green circle) and {\it XMM-Newton} (red circle) positions are also overlaid with error circles of $0\farcs67$ and $0\farcs80$ radius, respectively (90\% confidence, including the error on the calibration of the optical image). We also note the association of stars to the east of the ULX.}
   \label{hoixx1_errorcircle}
\end{figure*}

The {\it HST} images were individually astrometrically re-calibrated using 10 stars from the Two Micron All Sky Survey catalog \citep{2006AJ....131.1163S}. The rms of these fits are between $0\farcs14$ and $0\farcs18$. In the {\it Chandra} observations, the ULX is located almost on-axis, with an offset of only $\sim 30\arcsec$. In these conditions, the {\it Chandra} X-ray absolute position has a 90\% uncertainty of $0\farcs6$. We have compared the X-ray positions of some sources outside the field shown in Figure \ref{hoixx1_errorcircle} with the {\it HST}/ACS images and found that they are coincident with extended objects, hence preventing a re-calibration. Taking the optical centroids of these objects, we found that two of them are consistent with the {\it Chandra} X-ray position within $0\farcs15$ and the third one shows a larger offset, $0\farcs35$. The positions of the ULX in the three {\it HST}/ACS observations are all within 1 pixel, i.e., $0\farcs05$ and are totally consistent with the {\it Chandra} position (Table \ref{tab_positions_hoix}).

We also used for comparison the {\it XMM-Newton} position, Table~\ref{tab_positions_hoix}, coming from a long observation (PI: Tod Strohmayer, ObsID: 0200980101, exposure time: 100~ks) and listed in the second {\it XMM-Newton} serendipitous source catalog \citep{2009A&A...493..339W}. The XMM position has a statistical error of $0\farcs08$ but is affected by a $0\farcs35$ systematic uncertainty ($1\sigma$). Thus, the 90\% confidence radius, including the optical astrometric uncertainty, is $0\farcs8$. Figure~\ref{hoixx1_errorcircle} shows the {\it Chandra} and {\it XMM-Newton} positions of HoIX~X-1 with, respectively, a $0\farcs67$ and a $0\farcs80$ radius error circle (90\% confidence, including the error on the calibration of the optical image) on an $F435W$ {\it HST}/ACS image, which does not allow one to identify a unique counterpart. Specifically, the brightest object is contaminated by a nearby faint star, $V \sim 24.5$, at $\sim 0\farcs2$ northwest of its position.

\subsubsection{Spectroscopic identification}

Our SUBARU spectroscopic observations were centered on the presumed counterpart, i.e., the brightest object of the small association of stars. The ULX is embedded in the optical nebula MH9/10 but fortunately, the central position of the counterpart is coincident with a region where the nebula has low intensity.

\begin{figure*}[!th]
   \begin{tabular}{cc}
      \resizebox{8.5cm}{!}{\includegraphics{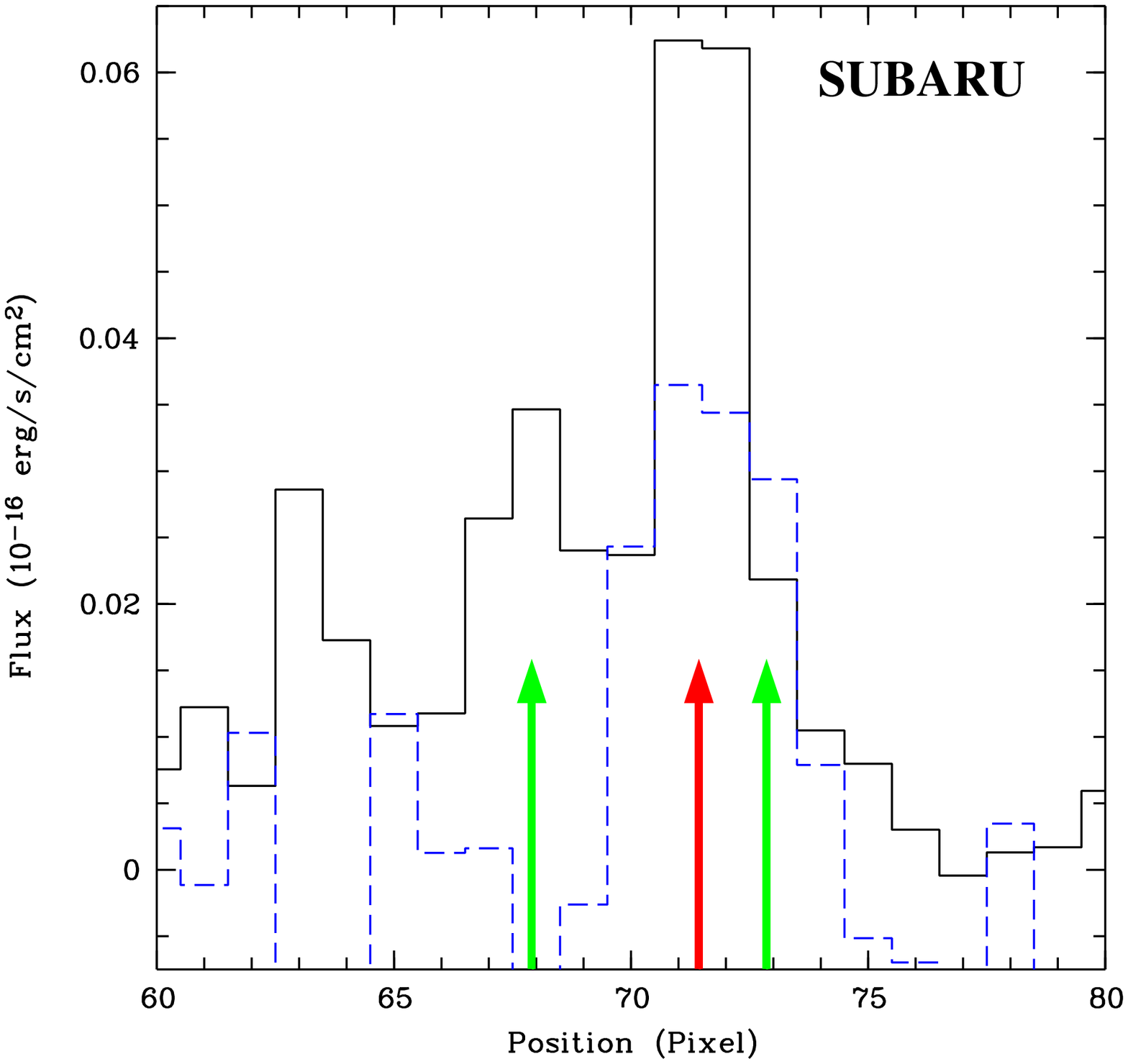}}
     &\resizebox{8.5cm}{!}{\includegraphics{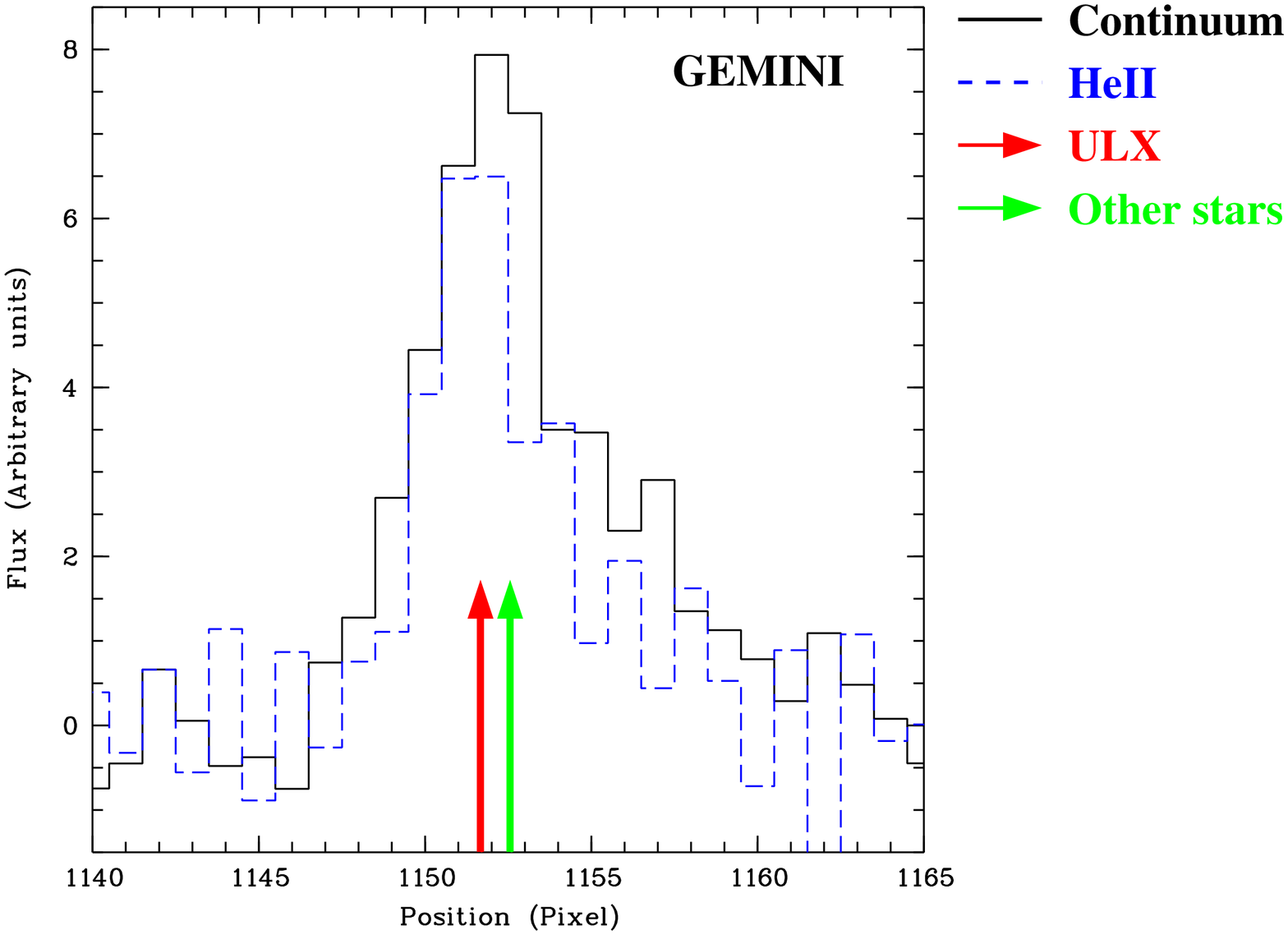}}\\
   \end{tabular}
   \caption[]{Spatial profiles of the continuum of the ULX (solid line) and of the \ion{He}{2} emission line (dashed line) in the two co-added spectra (left: SUBARU data, right: {\it GEMINI} data).
The continuum was taken between 4700 and 4714 \AA\ for the SUBARU data and averaged between 4662--4675 \AA\ and 4697--4711 \AA\ for the {\it GEMINI} data. The ULX is located near the peak at the pixel coordinate 71.5 (left) and near the pixel coordinate 1151.7 (right). The continuum of the ULX is slightly distorted, resulting from the presence of a nearby star at $0\farcs2$ at the northwest of its position (on the right here).
The vertical arrows show the position of the objects as derived from {\it HST} observations. In the SUBARU spectrum, we see that the peak of the \ion{He}{2} emission line is coincident with the brightest object in the {\it HST} images. In the {\it GEMINI} spectrum, the position angle for these observations was set to $180\degr$ ($90\degr$ for the SUBARU data). As can be seen in the {\it HST} image (Figure \ref{hoixx1_errorcircle}), three sources thus contribute to this spectrum. The tail of the emission at the left of the brightest object is due to a faint object located at $0\farcs6$ south of the main object, the tail at the right is due to the closest object discussed already. Finally, we can see that the peak of the \ion{He}{2} emission is again consistent with the position of the brightest object. This strongly suggests that the brightest object is the optical counterpart of the ULX.
}
   \label{heii_spatial}
\end{figure*}

The one-dimensional SUBARU spectrum of the most luminous object in the X-ray error circle with the slit oriented east-west is shown in Figure~\ref{spectres_1D_contrepartie_hoixx1}. The main feature of the spectrum is the presence of the $\lambda$4686 \ion{He}{2} emission line, which is a typical signature of X-ray binaries and thus identifies this object as the optical counterpart of the ULX. The line is spatially coincident with the continuum of the counterpart, Figure~\ref{heii_spatial}, and excludes the possibility that the emission comes from the nearby fainter star. The \ion{He}{2} line is not spectrally resolved in the FOCAS spectra. Its equivalent width (EW) is $12 \pm 3\ \mathrm{\AA}$. The emission coincident with the continuum of the counterpart is unresolved, Figure~\ref{heii_spatial}, as in NGC~1313~X-2 \citep{2006IAUS..230..293P,2009AIPC.1126..201G}, and thus probably reflects the emission of the ULX system.

\begin{figure*}[!t]
      \resizebox{7.5cm}{!}{\includegraphics{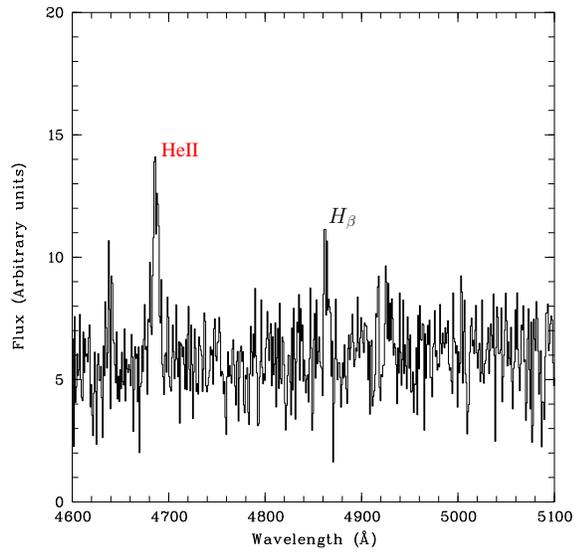}}
   \caption[]{4600--5100 \AA\ part of the {\it GEMINI/GMOS-N} one-dimensional spectrum of HoIX~X-1 optical counterpart, confirming the presence of the \ion{He}{2} line at 4686 \AA\ . The other annotated line in emission is from the nebula in which the counterpart is located and which result probably from our rough subtraction considering the highly variable profile of the nebula.}
   \label{spectres_1D_contrepartie_hoixx1_gemini}
\end{figure*}

The {\it GEMINI} spectrum confirms this result, see Figure~\ref{spectres_1D_contrepartie_hoixx1_gemini}. In this case, the slit was oriented north-south so the brightest object and nearby fainter star are located at almost the same position along the slit. The \ion{He}{2} line is again spatially coincident with the brightest object. Thus, the fainter object cannot be the ULX counterpart, in agreement with the SUBARU spectrum. The {\it GMOS} spectrum has better resolution (pixel scale of 0.9 versus 2.8 \AA\ $\mathrm{pixel^{-1}}$ and spectral resolution of 3.7 versus 8.0 \AA\ for SUBARU spectrum), allowing the emission lines to be resolved. In the co-added spectra, the \ion{He}{2} line has an intrinsic FWHM of 7.0 \AA\ or 450 $\mathrm{km\ s^{-1}}$ compared to the nebular lines like [\ion{O}{3}]$\lambda$5007 with an intrinsic FWHM $\sim 3.1$ \AA\ or 200 $\mathrm{km\ s^{-1}}$. The width of the \ion{He}{2} line is comparable with that of NGC~1313~X-2 \citep{2006IAUS..230..293P,2009AIPC.1126..201G} and indicates that the \ion{He}{2} emission is produced in the ULX system rather than being of nebular origin. Contrary to the behavior observed in NGC~1313~X-2 \citep{2009AIPC.1126..201G}, the EW of the \ion{He}{2} line is stable in HoIX~X-1 between the SUBARU and the {\it GEMINI} observations, with a value of 11 \AA\ in the latter which is consistent with the high and persistent X-ray luminosity of the ULX. We note that the EW of the \ion{He}{2}$\lambda$4686 line is more than 10 times higher in this object than in the brightest Galactic (and Magellanic) high-mass X-ray binaries which is consistent with the higher X-ray luminosity of ULXs.

The radial velocity of the \ion{He}{2} line is similar in the six individual {\it GEMINI} observations and does not show significant variation between the SUBARU and the {\it GEMINI} observations with a velocity consistent with those of the nebular lines. Although this result could just be a matter of coincidence, we would expect a radial velocity shift if the \ion{He}{2} line is indeed emitted in the accretion disk and the black hole is not very massive. We note that a recent dedicated spectroscopic campaign done by \citet{2010arXiv1011.2155R} has shown that the \ion{He}{2} line displays non-periodic radial velocity variations with rapid variability of the line profile on timescales shorter than a day, casting doubt on the possibility to derive a mass function for this object with this method.

Thus, the spectrum of the counterpart of HoIX~X-1 is similar to other spectra of ULX counterparts (e.g, \citealt{2009AIPC.1126..201G,2009ApJ...697..950K}), specifically a blue continuum without any spectral features apart from the \ion{He}{2} emission line. The two-dimensional spectrum of the counterpart is contaminated by the presence of nebular emission lines. As the intensity of the nebula is highly spatially variable, we extracted the spectrum by subtracting a nebular background taken from a small region close to the ULX counterpart on the faint side of the nebula. We note the presence of emission lines such as [\ion{O}{3}], [\ion{O}{1}], H$_\alpha$, and [\ion{S}{2}], which are probably residual emission of the nebula due to its highly non-uniform intensity.

\subsection{Stellar association around the ULX}
\label{ulxpop}

\begin{figure*}[t]
	\rotatebox{0}{\resizebox{15cm}{!}{\includegraphics{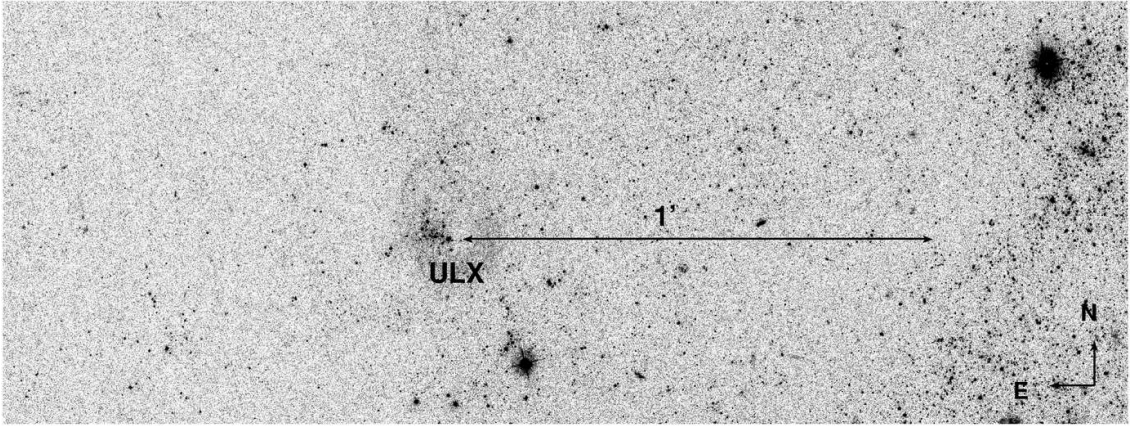}}}
   \caption[]{Location of the ULX compared with the denser star formation areas of Holmberg~IX, on the west edge, on an {\it HST}/ACS observation in the $F435W$ band.}
   \label{hoixx1_distance_zones_stellaires}
\end{figure*}

The multicolor SUBARU $BVR$ image (Figure~\ref{hoixx1_images_neb_couleur}) shows that the environment around the X-ray source is dominated by a compact group of blue stars located inside the H$_\alpha$ nebula, the $R$ filter also includes the H$_\alpha$ line. Outside of the nebula, we observe a mix of blue and red stars, and also foreground stars. On a large-scale {\it HST} image (Figure~\ref{hoixx1_distance_zones_stellaires}), we note that HoIX~X-1 is located about $1\arcmin$ east of an area rich in blue stars and is outside the standard $D_{25}$ isophotal radius of the galaxy Holmberg~IX. On the contrary, we observe few stars north of the ULX, with a diffuse background of red stars close to the limit of detection and a few blue stars.

\begin{figure*}[t]
   \begin{tabular}{cc}
      \includegraphics[width=9.5cm, angle=0]{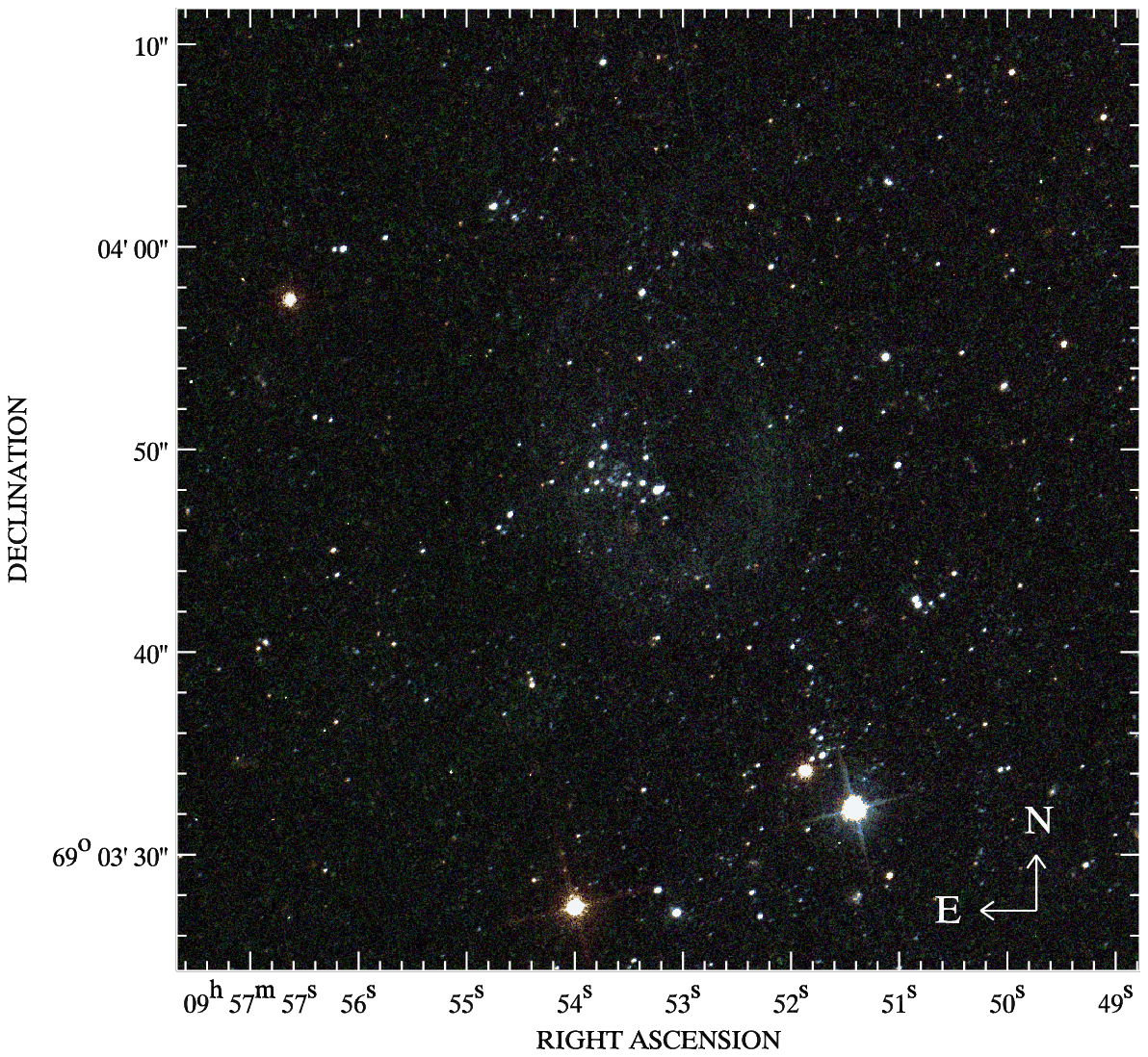}
     &\includegraphics[width=8.1cm, angle=90]{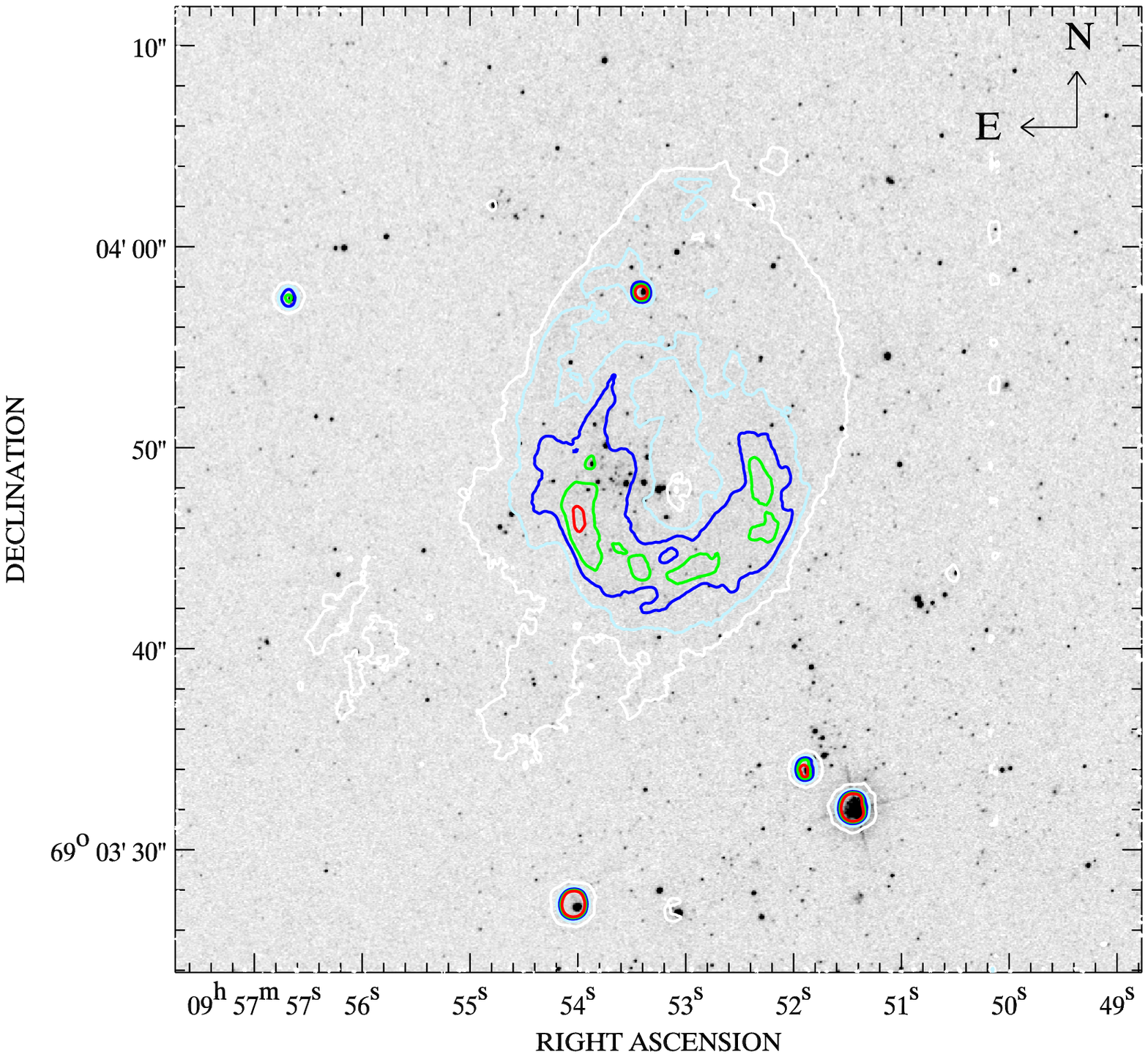}\\
   \end{tabular}
   \caption[]{Left: true color image (blue: $F435W$; green: $F555W$; red: $F814W$) 
of the region around HoIX~X-1, from {\it HST}/ACS observations. $1 \arcsec$ represents 17.4 pc at the distance of Holmberg~IX. 
The stellar environment is quite poor in the region close to the ULX.
We can mainly see a blue association localized to the east of the counterpart.
Right: same view of the vicinity of the counterpart, in the $F435W$ filter. 
The ULX counterpart is the brightest point-like source at the center of the image. Contours of the H$_{\rm{\alpha}}$ emission (at 10\%, 20\%, 40\%, 60\% and 80\% flux level above the background) are overplotted, from our SUBARU observations.
The stellar association is localized or projected inside the H$_{\alpha}$ nebula.
}
   \label{hoix_hst_bvi}
\end{figure*}

The ULX clearly does not belong to an intense star-forming region but rather appears outside the main body of the galaxy. The tight stellar association around the ULX is resolved into some seven stars (Figure \ref{hoixx1_images_neb_couleur}) in the SUBARU data thanks to the excellent seeing during the observations (Table \ref{tab_subarudata}). The multicolor {\it HST} image (Figure \ref{hoix_hst_bvi}) resolves the stellar association close to the ULX into dozens of members.

\subsection{Extinction}

\begin{figure*}
	\includegraphics{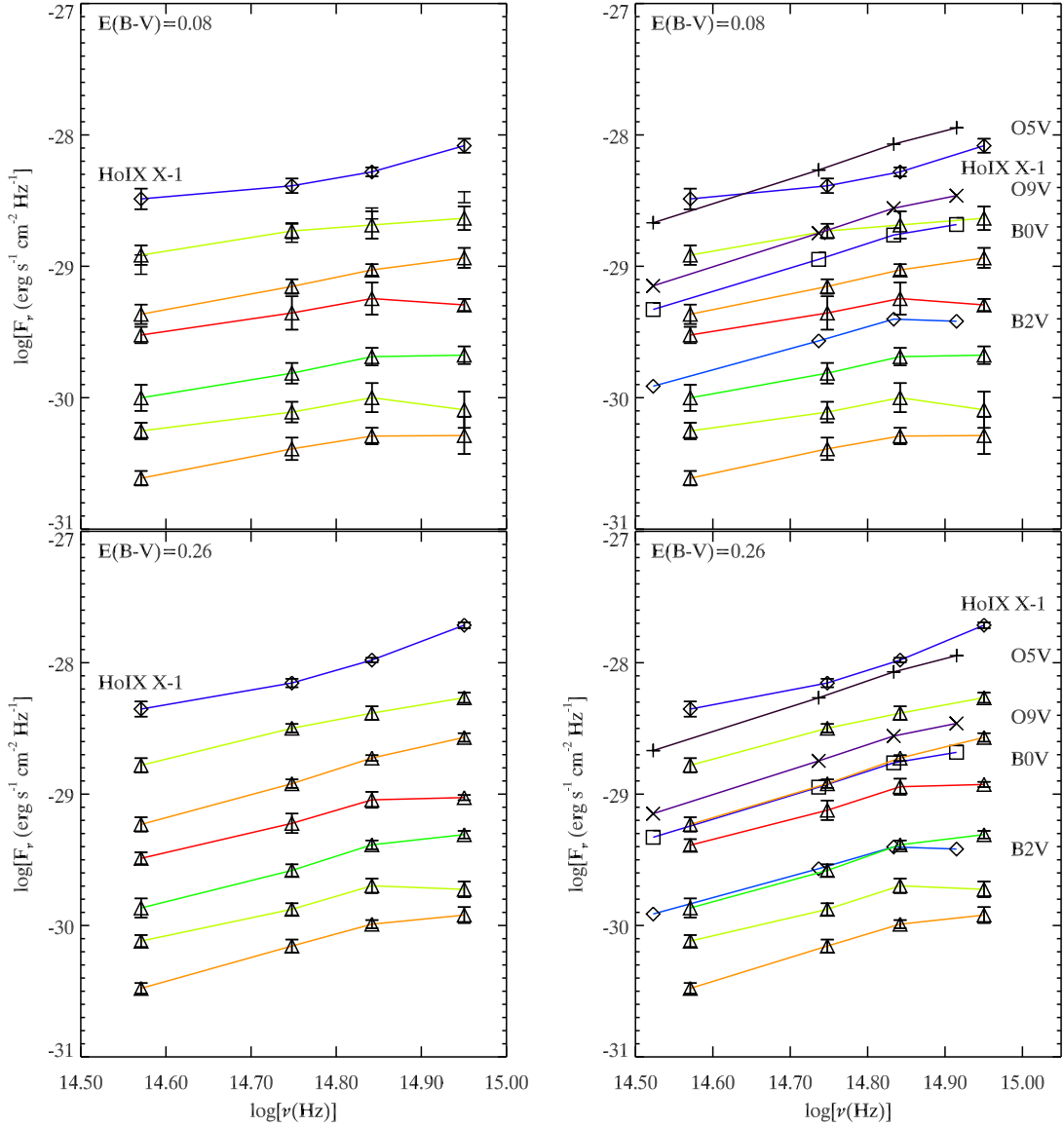}
   \caption{Spectral energy distribution (SED) for HoIX~X-1 and six of the following brightest stars that have been shifted down for clarity purpose. Upper panels: the use of the Galactic extinction $E(\bv)=0.08$ fails to reproduce the slope of the expected OB stars. Lower panels: increasing the extinction to $E(\bv)=0.26$ makes the SEDs of the stars fairly consistent with those of OB stars templates. Left panels show only the SEDs of the stars near the ULX (triangles) and right panels show in addition stellar templates designated by their spectral type.}
   \label{hoix_sed_compstars}
\end{figure*}

The quality of the photometry and the good agreement between the two data sets are two major points. Another key parameter for the interpretation of the data is the extinction along our line of sight toward HoIX~X-1. The Galactic reddening is low ($E(B-V)$ = 0.08; \citealt{1998ApJ...500..525S}), and studies of the stellar population in Holmberg~IX have never clearly identified any substantial, local reddening (e.g., \citealt{2002A&A...396..473M}). To check the extinction, we plotted the spectral energy distributions (SEDs) of all the stars located near the ULX dereddened by the Galactic extinction, see Figure~\ref{hoix_sed_compstars}. Given the magnitudes and colors of most of those stars ($\bv \sim 0.1$, $V-I \sim 0.0$, $-5< M_V < -2$), they have to be main-sequence early-type stars, such as O8 to B2 stars, although we cannot rule out that a few may be evolved. Comparison of these stars with SED templates of early-type stars \citep{Schmidt-Kaler1982,1992msp..book.....S,2000asqu.book.....C} show that their slope is not in agreement. The extinction must be increased to $E(\bv) \sim 0.3\ \mathrm{mag}$ to make the observed fluxes consistent with the SED templates, Figure~\ref{hoix_sed_compstars}. We note that this estimate is made using only the $BVI$ photometric bands (and close related {\it HST} bands $F435W$, $F555W$, $F814W$) because the $U$ band is different from the $F330W$ band, the former lying on top of the Balmer jump and the latter being located shortward of the jump. We also studied the variation of the Balmer decrement (H$_{\alpha}$/H$_{\beta}$) in the spectrum of the nebula and we found that MH9, the east side of the nebula (along the slit) shows a higher value compared to MH10 (the west part), i.e., $E(\bv) = 0.26 \pm 0.04\ \mathrm{mag}$ compared to $E(\bv) = 0.09 \pm 0.05\ \mathrm{mag}$. This suggests differential extinction in the nebula. In addition, the SEDs of several stars located in a small region west of the ULX indicate higher reddening, up to $E(\bv) \sim 0.8\ \mathrm{mag}$. In the remainder of this paper, we will use our best estimate of the extinction for the ULX, $E(\bv) = 0.26 \pm 0.04\ \mathrm{mag}$, with the caveat that there may be some variation of the extinction.

\subsection{Interpretation of the Color-Magnitude diagrams}

\begin{figure*}
\includegraphics{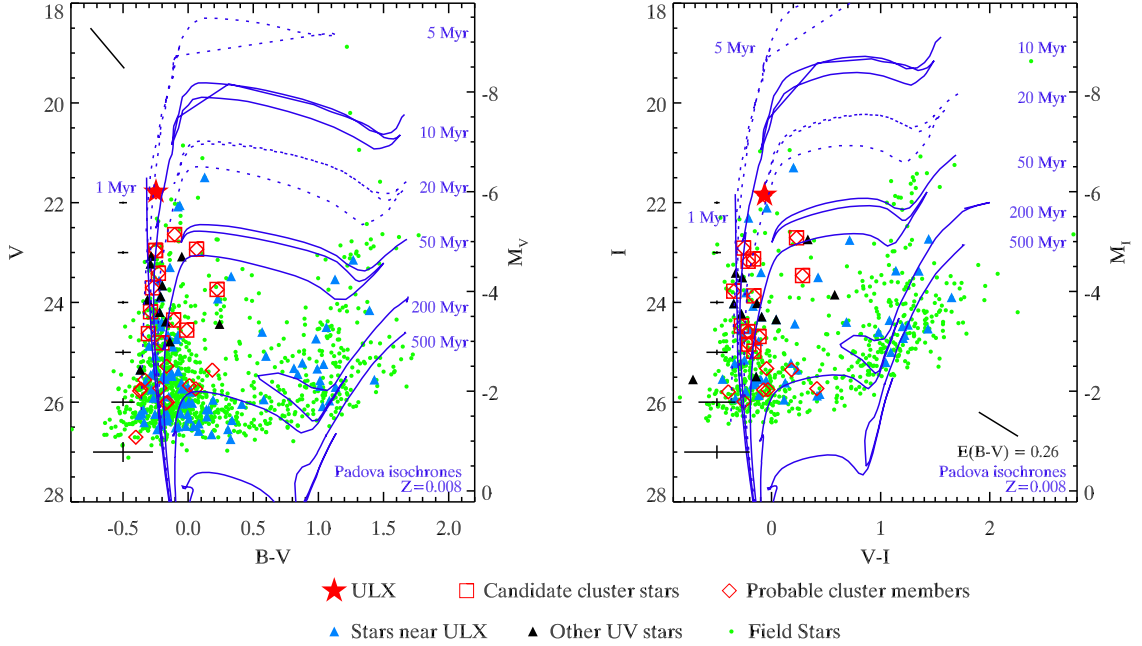}
   \caption[]{{\it HST}/ACS color-magnitude diagrams for the stellar field ($2000 \times 2000\ \mathrm{pixels}$, i.e., $100\arcsec \times 100 \arcsec$)
   around the ULX. {\it HST}/ACS magnitudes were transformed into the 
   Johnson-Cousins system. Padova isochrones for stars of different ages are 
   overplotted. Typical photometric errors are also plotted. Data 
   have been corrected for the derived extinction ($E(B-V)
   = 0.26$ mag), the bar at the top left corner illustrating this
   effect.
   ``Stars near ULX'' are stars not located inside the H$_{\alpha}$ nebula but within $25\arcsec$ of the ULX position. ``Candidate cluster stars'' are those located inside the nebula and detected in the UV image ($F330W$ filter). ``Probable cluster members'' are those inside the nebula but not detected in the UV image, with $\bv < 0.5$. The ``Other UV stars'' are stars detected in the UV but which are located outside the nebula. Finally, ``Field stars'' are objects which are not compatible with any of these criteria.
   Left panel: color-magnitude diagram in the ($B$,$V$) system.
   Right panel: color-magnitude diagram in the ($V$,$I$) system. The same
   isochrones are plotted in the two panels, i.e., for 1, 5, 10, 20, 50, 200, and
   500 Myr at $Z = 0.008$.} 
   \label{cmd_johnson}
\end{figure*}

Color-magnitude diagrams in $\bv$ and $V-I$ (Figure \ref{cmd_johnson}) show the predominance of a blue main sequence ($\bv \sim -0.2$, $V-I \sim -0.2$) but also the existence of a giant branch, especially obvious in the $V-I$ diagram, as pointed out by \citet{2008ApJ...676L.113S} with their observations centered on Holmberg~IX.  The population of stars apparently associated with the ULX is nicely resolved with {\it HST} and we can see 10 members well identified by their positions internal to the nebula. The UV emission as seen in the $F330W$ filter, see Figure~\ref{hoixx1_errorcircle}, provides a further clue to their relatively young age. In our study of NGC~1313~X-2 (G08), comparison with the $F330W$ image allowed us to discriminate between young stars and old field stars in a semi-quantitative manner. Here, the situation is more complex since the dominant population is young, likely because the encounter between M81, M82, and NGC~3077 triggered a wave of star formation within the past 200 Myr. Nevertheless, we can assign the status of ``candidates of the association'' to all stars visible in the UV image and which are visually in the dense association.

This group of stars does not differ from the other field stars but fits very well in the main sequence defined by the other stars. Figure~\ref{cmd_johnson} shows a color-magnitude diagram corrected by the assumed reddening $E(\bv) = 0.26$ and an extinction based on the Cardelli law \citep{1989ApJ...345..245C}. The Padova isochrones \citep{2008A&A...482..883M,2010ApJ...724.1030G} used are for a metallicity $Z=0.008$, as derived by \citet{2002A&A...396..473M} by spectroscopy of an \ion{H}{2} region (region 8 from \citealt{1994ApJ...427..656M}; located $2.5\arcmin$ from HoIX~X-1) and consistent with the dwarf nature of the galaxy. Concerning the age determination, we are facing a situation where most of the stars are located near the 1--20 Myr isochrones on the main sequence. The width of the main sequence, explained by photometric errors and probably some differential reddening, is quite large compared to the separation of the 1--20 Myr isochrones. We note that there are three stars (five if we consider the probable two reddened stars) with an absolute $V$ magnitude between -4.5 and -5.0. These stars may be late-O type stars or early-B stars evolving toward the supergiant phase. Thus, the age of this stellar association is uncertain but is most likely younger than 20 Myr.

We note that there is a population of stars in the field that is probably older, $\sim 50\ \mathrm{Myr}$ old, as seen by the $\sim 10$ red supergiants at ($M_{V} \sim -5$; $\bv \sim 1.4$, $V-I \sim 1.3$). There are no such red supergiants associated with the stellar association around the ULX, but as we remarked in G08, the ratio of blue to red supergiants observed in metal-poor galaxies \citep{1995A&A...295..685L} being $\sim 3$ would imply only one or two red supergiants in the association which, given the statistics, is not in contradiction.

A previous study done by \citet{2006ApJ...641..241R} found an age of 4--6 Myr using the same {\it HST} data in the $B$ and $V$ filters. They in fact assumed that the blue stars in the color-magnitude diagram are reddened ($E(\bv)=0.2$) main-sequence O stars. Even though this assumption had no real basis, it seems that our finding, based on a more careful extinction estimate, may be consistent with a close value.

\begin{figure*}
\includegraphics{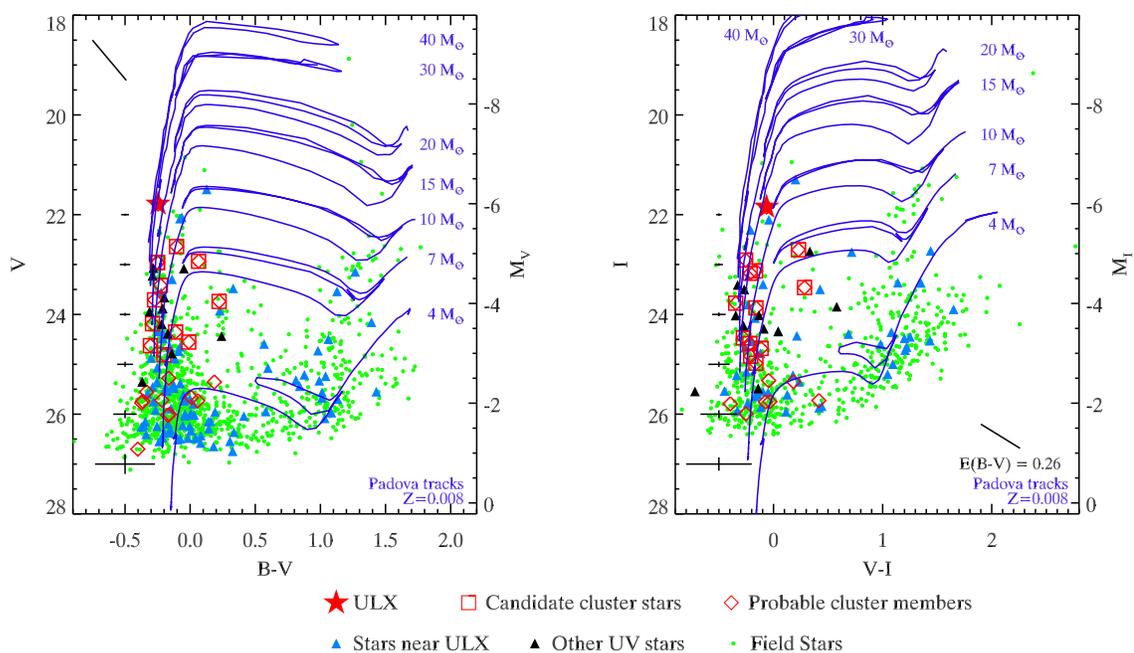}
   \caption[]{{\it HST}/ACS color-magnitude diagrams (Johnson-Cousins photometry)
   with Padova evolutionary tracks for stars of different initial masses
   ($4$, $7$, $10$, $15$, $20$, $30$, and $40\ \mathrm{M_{\sun}}$) with $Z=0.008$.\
   The legend is the same as Figure \ref{cmd_johnson}.
  Left panel: color-magnitude
   diagram in the $(B,V)$ system. Right panel: color-magnitude diagram in
   the $(V,I)$ system.} 
   \label{cmd_ET}
\end{figure*}

In terms of mass, the Padova evolutionary tracks show that the most luminous stars are compatible with a zero-age mass of the order of 10--20 $M_{\odot}$ (Figure \ref{cmd_ET}). We note also that in principle, the $F330W$ band would be more discriminating in terms of age and mass of the stellar population, but unfortunately there is no available observational calibration between the $F330W$ and the more common $U$ band \citep{2005PASP..117.1049S}. Finally, using a solar metallicity ($Z=0.019$) instead of the sub-solar one assumed here ($Z=0.008$) does not influence our results in any significant manner.

\subsection{Properties of the stellar association}

The association of stars around HoIX~X-1 looks more like an OB association than a cluster linked by gravitation. This is something we also saw in NGC~1313~X-2 (G08). The integrated magnitudes of the stellar association are $M_B \sim -7.6$, $M_V \sim -7.4$ and $M_I \sim -7.3$. Comparing these magnitudes with Starburst99 simulations \citep{1999ApJS..123....3L}, and assuming that the age of the cluster is 10/20 Myr with a Salpeter initial mass function (IMF) or an ``$\alpha = 3.30$'' IMF (with masses between $1$ and $100\  M_{\odot}$), the mass of the cluster is about $M=(1.5 \pm 0.3)/(2.7 \pm 0.3) \times 10^{3} M_{\odot}$, which is similar to the association surrounding the ULX of our previous study (G08). This group of stars is clearly typical for OB associations of the Local Group, both in terms of mass and of size, with its diameter of 90 pc ($\sim 5\arcsec$). This isolated star formation is probably linked to the tidal forces between M81 and M82 which have created many stars concentrations, looking like either dwarf galaxies (like Holmberg~IX) or less massive associations as recently discovered by \citet{2008AJ....135..548D} and \citet{2008PASP..120.1145D}.

\section{Nature of the ULX optical counterpart}

\subsection{Constraints from photometry}

\begin{figure*}
      \resizebox{10cm}{!}{\includegraphics{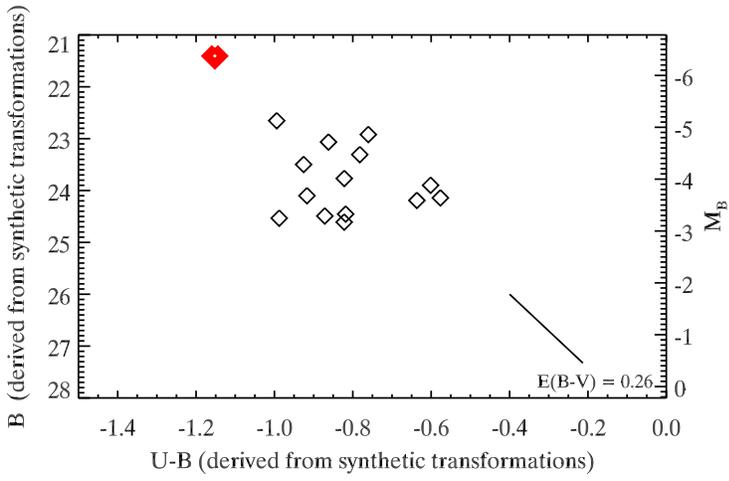}}
   \caption{{\it HST}/ACS color-magnitude diagram for the stellar field around the ULX in the synthetic ($U$,$B$) system. The ULX counterpart (red diamond) is clearly bluer than the other stars of the association, contrary to the other color-magnitude diagrams.}
   \label{cmd_ub_hoix}
\end{figure*}

We have shown that the stellar population around the ULX has an age of $\la 20\ \mathrm{Myr}$ and an individual upper mass limit of $15^{+5}_{-5}\ M_{\mathrm{\odot}}$. In contrast to the NGC~1313~X-2 (G08), the ULX of our present study has a counterpart brighter than the other stars belonging to the stellar association, by almost 1 mag. If the counterpart is a normal star (as opposed to the case where the accretion disk may play a predominant role in the optical emission), its magnitude $M_V \sim -6.0$ would be consistent with an O3--O4 star on the main sequence, a late-O type evolved star, or a late-O/early-B Ib supergiant. Its dereddened color $(B-V)_0 \sim -0.25$ (Table \ref{tab_mag_colors}) would be consistent with most of those scenarii, but its $(V-I)_0 \sim -0.06$ is somewhat too red to be compatible with such interpretation. Indeed, the counterpart shows a slightly ``red'' excess (compared to the other stars of the association) since its ($V-I$) color would be more typical of a late-B star on the main-sequence, or a mid supergiant B star. But perhaps the most important feature we see is a significant excess in the UV (compared to the other stars of the association). This can be seen in the color-magnitude diagram ($U-B$) versus $B$, Figure~\ref{cmd_ub_hoix}, where the ULX counterpart is clearly bluer than any other star of the stellar association. Unfortunately, this diagram is only illustrative of this UV excess since the magnitudes and colors $U$, $B$ and ($U-B$) derived from {\it HST/ACS} data are not compatible with those used in the available evolutionary tracks and isochrones and thus cannot be used to derive age or mass properties of the stars.

\begin{figure*}
     \includegraphics{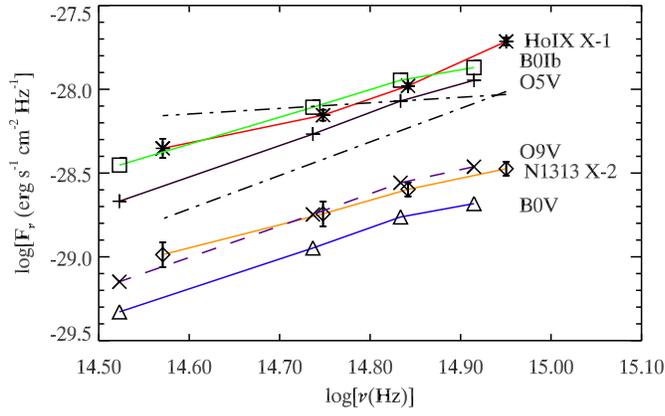}
   \caption{Spectral energy distribution (SED) for two ULXs including the object of our study, with the SEDs of different OB star templates. The two dot dashed black lines are power laws with indices of 2.0 and 1/3, respectively corresponding to a Rayleigh-Jeans tail and the expected spectrum of a thin accretion disk.}
   \label{sed_hoix}
\end{figure*}

The SED of the ULX counterpart, Figure~\ref{sed_hoix}, leads to the same conclusion.  Compared to other stars of the association, including the brightest ones in the UV, we can definitely see that the ULX counterpart shows a clear excess in the UV and a slight excess in the red part of the spectrum.  In fact, it shows a 3363--4317 $\mathrm{\AA}$ slope steeper than a O5 star, pointing to a very hot continuum in the UV. Figure~\ref{sed_hoix} also shows the SED of the ULX NGC~1313~X-2 (G08); NGC~1313~X-2 looks like a normal OB star which is relatively unexpected since we have shown in G08 that the optical variability of the source was likely associated with some X-ray reprocessing in the system. This deserves more detailed study. Interestingly, \citet{2009ApJ...697..950K} showed that the ULX NGC~5408~X-1 also displays a UV excess at wavelengths shorter than the Balmer edge. They interpret this effect as coming from the accretion disk and the photoionized nebula surrounding the counterpart. In our case, the main nature of the nebula is quite different since it seems to be mainly shock-ionized \citep{2006IAUS..230..302G,2007AstBu..62...36A,2008AIPC.1010..303P}, but X-ray/UV photoionization is possibly required to explain all the properties of the ULX nebula \citep{2008RMxAA..44..301A}. Also, Roberts et al. (private communication) have shown, using very deep optical spectra of these two ULXs, that NGC~1313~X-2 may display some absorption features coming from the donor star (specifically, a Balmer jump), while HoIX~X-1 shows only a very blue continuum consistent with our findings. In any case, the absence of clear absorption lines in both ULXs point to a contribution from the accretion disk, which is probably acting at different levels and/or at different wavelengths in both sources.

The X-ray to optical flux ratio, $\xi = B_0 + 2.5\ \mathrm{log} F_X$ (e.g., \citealt{2005ApJ...629..233K}), with $F_X$ the 2--10 keV unabsorbed X-ray flux density in $\mu Jy$ and $B_0$ the dereddened $B$ magnitude, is $\xi = 20.6$. This is consistent with the values found in LMXBs \citep{1995xrbi.nasa...58V} where the optical emission is dominated by X-rays reprocessing. We remark that if the accretion disk and the secondary star had equal luminosities, subtracting the light from the accretion disk would correct the position of the counterpart to be consistent with a $M_V \sim -5.3\ \mathrm{mag}$ star. Thus, it would simply be as luminous as the brightest stars in the tight association. The UV excess, as in low-mass X-ray binaries (LMXBs), would be due to reprocessed energy radiated in the UV and optical \citep{1995xrbi.nasa...58V}. A key problem in ULXs is that we do not know the spectral type of the companion star.

A few models predicting the optical properties of ULXs have appeared recently in the literature \citep{2005MNRAS.356..401R,2005MNRAS.362...79C,2007MNRAS.376.1407C,2008ApJ...688.1235M,2008MNRAS.386..543P}. Although their conclusions are not unanimous, one common result of these studies is that the companion star should be massive enough, $M > 10\ M_{\mathrm{\odot}}$, to sustain a stable accretion rate and to allow for X-ray luminosities of $10^{40}\ \mathrm{erg/s}$. This was also pointed out in \citet{2003MNRAS.341..385P} where the authors show that steady accretion can be achieved with a donor star of $8$--$15\ M_{\mathrm{\odot}}$ around a $10\ M_{\mathrm{\odot}}$ black hole and can lead to X-ray luminosities above $10^{40}\ \mathrm{erg\ s^{-1}}$ during $9$--$5\times 10^{5}\ \mathrm{yr}$, respectively. These phases of high X-ray luminosity happen when the donor becomes a giant and the evolution is driven by the nuclear evolution of the hydrogen-burning shell \citep{2003MNRAS.341..385P}. Observational results also suggest giant companions for some other ULXs (e.g., \citealt{2004ApJ...602..249L,2006ApJ...646..174K,2008ApJ...675.1067F,2010arXiv1011.4215M}).

The most recent published model predicting the optical output of ULXs \citep{2008MNRAS.386..543P,2010MNRAS.403L..69P} is also the only one that takes into account the fact that not all the mass lost by the donor star is accreted by the black hole. Instead, and especially when the donor star is leaving the main sequence, the rise in the mass transfer rate above the Eddington limit implies that there is substantial mass loss from the binary that does not contribute to the X-ray luminosity and to the reprocessing \citep{2008MNRAS.386..543P,2010MNRAS.403L..69P}. One of the main differences with previous models is that accretion disks cannot be very bright in the optical unless the black hole is massive enough and/or the donor star is in the fast expanding giant phase.

If we apply our photometric results to this model, HoIX~X-1 is consistent with a $30$--$50\ M_{\mathrm{\odot}}$ donor around a $20$--$100\ M_{\mathrm{\odot}}$ black hole undergoing case A(B) mass transfer. The donor is likely to be ascending the giant branch, which would give rise to a larger mass accretion rate and thus, a larger X-ray luminosity. However, this stage is likely to be very short for such a massive star, only a few tens of thousands of years \citep{2008MNRAS.386..543P}. Based on the same arguments applied to NGC~1313~X-2 by \citet{2010MNRAS.403L..69P}, case B mass transfer is unlikely due to the fact that the ULX is surrounded by a nebula that is $\sim 1\ \mathrm{Myr}$ old \citep{2008AIPC.1010..303P,2008RMxAA..44..301A}.  Thus, case AB, where mass transfer starts during the main sequence, is more likely since it allows several millions years \citep{2008MNRAS.386..543P} to inject the required energy to the nebula. In that framework, our values of $V$ magnitude and $\bv$ color do not allow discrimination of the mass of the black hole. Addition of other photometric bands in binary models may help in further constraining the mass of the components. Obtaining a full SED of the ULX from the far-UV to the near-infrared would also be helpful in trying to disentangle the contributions coming from the accretion disk versus the donor star.

From our own results, the firm conclusion that we can draw about the companion star is that it is likely a quite massive star, $M \ga 10\ M_{\mathrm{\odot}}$, when we take its position in the color-magnitude diagrams as those of a non-irradiated star. The same conclusion holds if we take into account a likely contamination of the optical flux by the accretion disk. Unfortunately, applied to the \citet{2008MNRAS.386..543P} model, our results do not favor either the ``standard'' stellar-mass black hole ($\sim 10$--$20\ M_{\odot}$) or the more massive ($\sim 100\ M_{\odot}$) black hole scenario.  However, we showed that the total stellar-mass in the young stellar association around the ULX is $< 10^4 M_{\odot}$.  Thus, HoIX~X-1 is clearly not in a super-star-cluster. This means that whatever the nature of the compact object, it was not formed via a runaway-coalescence scenario as proposed by \citet{2001ApJ...562L..19E} and \citet{2002ApJ...576..899P}. If the black hole has a mass of the order of $\la 100\ M_{\mathrm{\odot}}$, it was probably formed via a more conventional stellar evolutionary scenario. This may be possible thanks to the sub-solar metallicity of the ULX environment, see \citealt{2009MNRAS.400..677Z} for such arguments.

\subsection{Photometric variability}

We have tried to characterize the possible optical variability of the counterpart. We have four SUBARU images in the $V$ band, but only two of them have a good enough seeing to separate, via PSF fitting, the luminous counterpart from the fainter object which tends to contaminate the optical emission. In these two images, the counterpart has a constant $V$ magnitude within the photometric errors ($22.71 \pm 0.04$, $22.67 \pm 0.04$).

On the contrary, the two {\it HST}/ACS images in the $V$ band show variability, $\Delta V = 0.136 \pm 0.027$ clearly significant at $5 \sigma$. The amplitude is similar to that of NGC~1313~X-2 (G08). This optical variability may originate from ellipsoidal modulation reflecting the orbital movement of the donor star, like in massive X-ray binaries. It may also come from reprocessing of variable X-ray emission by the accretion disk and X-ray heating of the secondary as observed in LMXBs. A combination of both effects is likely, since the ULX environment is composed of moderately massive stars with $M_{\mathrm{max}} \sim 20 M_{\mathrm{\odot}}$, so that the optical emission of the donor is not negligible as in LMXBs, while the ULX X-ray output is 10--100 times that of LMXBs, suggesting a non-negligible optical contribution from reprocessing.

The SUBARU data do not permit a study without ambiguity of the intrinsic variability of the ULX, but the magnitude $V$ ($22.71 \pm 0.038$) from the 2003 January 26 observation is consistent with the range seen by {\it HST}. The $\bv$ color is consistent between the two observations of the two telescopes (Tables \ref{tab_magnitudes_subaru} and \ref{tab_magnitudes_hst_hoix}). We can compare the total magnitude of the counterpart and of the nearby object between the two observatories for all observations (Table \ref{tab_magnitudes_hoix_variabilite}). The agreement is very good between our SUBARU measurements (where we can resolve the counterpart from its nearby companion) and the {\it HST} observations. The second SUBARU observation (2003 January 26) does not allow the measurement of the faint object even if the measured total magnitude ($V=22.68 \pm 0.015$) suggests that we only see the ULX counterpart. The conclusion on the variability in the SUBARU observations is thus uncertain because of the photometric contamination by this fainter object. Nonetheless, all values are completely consistent with the luminosity range seen by the {\it HST}. We may be seeing similar variability as with the {\it HST}, but it is masked by the large uncertainties.

An interesting characteristic of these {\it HST} observations is that they were coordinated with two simultaneous {\it Chandra} ones. Unfortunately, the observations are highly piled-up preventing a detailed analysis. The derived fluxes suffer from a high uncertainty, with $F_X = 1.02^{+0.21}_{-0.15} \times 10^{-11}$, $1.00^{+0.21}_{-0.20} \times 10^{-11} \mathrm{erg\ s^{-1}\ cm^{-2}}$ for the two observations.\\
The optical variability observed $\Delta V \sim 0.14\ \mathrm{mag}$ corresponds to an increase of 14\% in the optical flux. Considering the errors in the X-ray fluxes, it is obvious that nothing can be said about a possible correlation between the X-ray and optical fluxes. Detailed studies are needed to establish X-ray/optical correlations and it is clear that no conclusion can be drawn from only two data points. But the fact that X-ray and optical variations are seen in other ULXs which do not seem to be entirely related to the orbital movement of the companion star \citep{2008A&A...486..151G,2009AIPC.1126..201G} are a strong motivation for future simultaneous X-ray/optical studies and will help to disentangle the mechanisms at the origin of the X-ray and optical emission.

\section{Conclusion}

We have used the high sensitivity of SUBARU/FOCAS and {\it GEMIMI/GMOS-N} and the high spatial resolution of {\it HST}/ACS to make a detailed study of the optical counterpart of the ULX HoIX~X-1 and its immediate environment. Close to the ULX, we have highlighted a group of young stars spread out over 100 pc that appears to be an OB association rather than a gravitationally linked cluster. The photometric properties of these stars are consistent with the young population that surrounds them. This tight association is quite isolated, located at 1 kpc from the dense star-forming area of Holmberg IX. This recent and localized episode  of star formation is possibly related to the past interaction between M81 and M82, similar to Holmberg IX and other structures formed at \ion{H}{1} overdensities. We have estimated that the young stellar association linked to the ULX has an age $\la 20\ \mathrm{Myr}$, a stellar-mass of $\sim 1$--3$\times 10^3\ M_{\odot}$, and an upper mass limit for single stars of about $20\ M_{\mathrm{\odot}}$. 
We have newly estimated the reddening value near the association and found a probable local extinction of $E(\bv) \sim 0.18\ \mathrm{mag}$ in addition to the very modest Galactic reddening of $E(\bv) \sim 0.08\ \mathrm{mag}$.

The optical counterpart of the ULX was identified by the presence of a broad, high-excitation \ion{He}{2}$\lambda$4686 line that testifies to X-ray reprocessing in the system. The counterpart appears to be the most luminous object in the association. Using stellar evolutionary tracks, we have constrained the mass of the companion star to be $\ga 10\ M_{\odot}$, even if the accretion disk contributes significantly to the optical luminosity. HoIX~X-1 displays a UV photometric excess compared to the surrounding population of stars. Its colors are not consistent with a star, and this prevents any spectral classification of the counterpart. The optical spectrum shows a blue continuum, consistent with the colors, with only high-excitation emission lines, similar to X-ray active LMXBs. Comparison of our photometric results with binary models suggest that the donor star may be as massive as $\sim 30$--$50\ M_{\mathrm{\odot}}$.

Another interesting result of this work is the variability of the ULX counterpart, by $\sim 0.15\ \mathrm{mag}$, clearly detected in the {\it HST}/ACS data set. However, we have insufficient information to determine if this variability comes from ellipsoidal variations, stochastically variable irradiation of the accretion disk or donor star, or some combination.

Finally, although no stringent constraints can be put on the mass of the black hole from these data, we can, as for the ULX NGC~1313~X-2 (G08), emphasize that the formation of an intermediate-mass black hole ($M > 100\ M_{\odot}$) in the environment of HoIX~X-1 is very unlikely, given the stellar-mass of the association and its unbound nature. This is consistent with recent X-ray results showing that this source can be modeled by what appears to be super-Eddington emission, hence a small ($< 100\ M_{\odot}$) black hole.

\clearpage

\begin{figure*}[!t]
   \begin{tabular}{cc}
      \resizebox{8cm}{!}{\includegraphics{f1}}
&      \resizebox{8cm}{!}{\includegraphics{f2}}\\
   \end{tabular}
   \caption[]{SUBARU/FOCAS composite images centered on the optical counterpart of the ULX.
Left: image $R$ (red) + $V$ (green) + $B$ (blue) ; right: image made of the nebular emission (H$_\alpha$-continuum -- red, [\ion{O}{3}]-continuum -- green) on which is superimposed the stellar emission coming from the $V$ band (blue).}
   \label{hoixx1_images_neb_couleur}
\end{figure*}

\clearpage

\begin{table*}
\caption[]{The SUBARU/FOCAS Observations for HoIX~X-1}
\label{tab_subarudata}
\centering
\small
\begin{tabular}{lcccc}
\hline
\hline
Filter & Exposure Time & Number of Exposures & Observation Date & Seeing \\
	& (s)		&			&		& ($\arcsec$)\\

\hline
$B$      &  300          & 2		     &	2003 Jan 25/26		 & 0.73/0.44		\\
$V$      &  300          & 4		     &	2003 Jan 25/26/26/Feb 09		 & 0.73/0.42/0.48/0.43	\\
$R$      &  300          & 2		     &	2003 Jan 25/26		 & 0.73/0.39		\\
H$_{\alpha}$ &  900    & 1		     &	2003 Jan 26		 & 0.51		\\
\ion{O}{3}   &  900    & 1		     &	2003 Jan 26		 & 0.53		\\
\hline
\hline
Spectrum & 1800/900    & 2		     &	2003 Jan 26			& 0.5		\\
\hline

\end{tabular}
\normalsize
\end{table*}

\clearpage

\clearpage

\begin{table*}
\caption[]{The {\it HST}/ACS Observations for HoIX~X-1}
\label{tab_hstdata}
\centering
\begin{tabular}{cccccccc}
\hline
\hline
ID & Instrument & Filter & Date & MJD at Mid-exposure  & Exposure Time (s)\\
\hline
j8ola2010 & HRC &  $F330W$  & 2004 Feb 7	& 53042.51347	 &     2760  \\
j8ol02040 & WFC &  $F435W$  & 2004 Feb 7	& 53042.46162	 &     2520 \\
j8ol02030 & WFC &  $F555W$  & 2004 Feb 7	& 53042.42239    &     1160  \\
j8ol02010 & WFC &  $F814W$  & 2004 Feb 7	& 53042.41364	 &     1160  	 \\
j8ol06010 & WFC &  $F555W$  & 2004 Mar 25	& 53089.45604	 &     2400   \\
\hline

\end{tabular}
\end{table*}

\clearpage

\clearpage

\begin{table*}
\caption{Luminosity of the ULX Counterpart in Different Filters from Our SUBARU/FOCAS Observations}
\centering
\begin{tabular}{ccccc}
\hline
\hline
Filter & Exposure Time (s) & Date     & Magnitude	& Absolute Magnitude\\
\hline
$B$      &     300      & 2003 Jan 26      & 22.72 $\pm$ 0.04 & -6.1\\
$V$      &     300      & 2003 Jan 26      & 22.71 $\pm$ 0.04 & -5.9\\
$R$      &     300      & 2003 Jan 26      & 22.64 $\pm$ 0.02 & -5.8\\
\hline

\end{tabular}
\label{tab_magnitudes_subaru}
\end{table*}


\begin{table*}
\begin{threeparttable}[b]
\caption[]{Brightness of the ULX Counterpart in Different Filters, from the {\it HST}/ACS Observations}
\label{tab_magnitudes_hst_hoix}
\centering
\small
\begin{tabular}{ccccccc}
\hline
\hline
Filter & Exposure Time & Date 	       & Aperture Correction & VEGAmag & Johnson Magnitude\\
\hline
$F330W / U$      &     2760 s  & 2004 Feb 7      & 0.790	& 20.833 $\pm$ 0.023   &   ...	 	 \\
$F435W / B$      &     2520 s  & 2004 Feb 7      & 0.427	& 22.541 $\pm$ 0.015   &   22.604	 \\
$F555W / V$      &     1160 s  & 2004 Feb 7      & 0.403	& 22.632 $\pm$ 0.024   &   22.609	  \\
$F814W / I$      &     1160 s  & 2004 Feb 7      & 0.716	& 22.334 $\pm$ 0.034   &   22.328 	  \\
$F555W / V$      &     2400 s  & 2004 Mar 25      & 0.384	& 22.768 $\pm$ 0.013   &   $\approx 22.75$     \\
\hline
\end{tabular}
\normalsize
\begin{tablenotes}
\item {\bf Note.} Magnitudes are expressed both in the {\it HST}/ACS Vegamag system and in the Johnson-Cousins ({\it UBVRI}) system, when possible.
\end{tablenotes}
\end{threeparttable}
\end{table*}

\clearpage

\begin{table*}
\caption[]{The {\it Chandra} Observations for HoIX~X-1}
\label{tab_chandradata}
\centering
\small
\begin{tabular}{ccccccc}
\hline
\hline
ID   & MJD  & Exposure Time & $n_\mathrm{H}$	& $\Gamma$	& $F_{X_{0.3-10\ \mathrm{keV}}}$	& Cash Stat./dof \\
     &	Mid-exposure		  &      (ks)	  & $(\mathrm{10^{21}\ cm^{-2}})$ &	&  ($10^{-11}\ \mathrm{erg\ s^{-1}\ cm^{-2}}$) &	\\	
\hline
4751 & 53042.45496	 &  5	& 1.0	& $1.79^{+0.19}_{-0.19}$	&  $1.11^{+0.17}_{-0.19}$ & 5.8/9  \\
4752 & 53089.41938	 &  5 	& 1.0   & $1.74^{+0.19}_{-0.19}$	&  $1.15^{+0.30}_{-0.21}$ & 6.6/7  \\
\hline

\end{tabular}
\normalsize
\end{table*}


\begin{table*}
\caption[]{Position of the ULX, Derived from the {\it Chandra} and {\it HST} Data Sets and Compared with the {\it XMM-Newton} Position}
\label{tab_positions_hoix}
\centering
\small
\begin{tabular}{ccc}
\hline
\hline
Telescope/Instrument & Right Ascension & Declination \\
\hline
{\it Chandra}            &    $09^{\mathrm{h}} 57^{\mathrm{m}} 53\fs32$ & $69\degr 03\arcmin 48\farcs10$ \\
{\it HST}/ACS $F435W$      &    $09^{\mathrm{h}} 57^{\mathrm{m}} 53\fs31$ & $69\degr 03\arcmin 48\farcs08$\\
{\it HST}/ACS $F555W$      &    $09^{\mathrm{h}} 57^{\mathrm{m}} 53\fs31$ & $69\degr 03\arcmin 48\farcs05$\\
{\it HST}/ACS $F814W$      &    $09^{\mathrm{h}} 57^{\mathrm{m}} 53\fs31$ & $69\degr 03\arcmin 48\farcs10$\\
{\it XMM-Newton}         &    $09^{\mathrm{h}} 57^{\mathrm{m}} 53\fs20$ & $69\degr 03\arcmin 47\farcs7$\\
\hline
\end{tabular}
\normalsize
\end{table*}

\clearpage

\clearpage

\begin{table*}
\begin{threeparttable}[b]
\caption[]{Observed and Dereddened Magnitudes and Colors of the ULX Counterpart, from the {\it HST}/ACS Observations}
\label{tab_mag_colors}
\centering
\small
\begin{tabular}{ccccc}
\hline
\hline
Filter/Magnitude & Observed Magnitude/Color & Dereddened Magnitude/Color\\
\hline
$B$      &     22.604 $\pm$ 0.015  & 21.536 $\pm$ 0.015	 \\
$V$      &     22.609 $\pm$ 0.024  & 21.780 $\pm$ 0.024	  \\
$I$      &     22.328 $\pm$ 0.034  & 21.844 $\pm$ 0.034 	  \\
\bv    &     -0.005 $\pm$ 0.028  & -0.247 $\pm$ 0.049	  \\ 
$V$--$I$    &0.281  $\pm$ 0.042  & -0.064 $\pm$ 0.058     \\
\hline
\end{tabular}
\normalsize
\begin{tablenotes}
\item {\bf Note.} Values are expressed in the Johnson-Cousins ({\it UBVRI}) system.
\end{tablenotes}
\end{threeparttable}
\end{table*}

\clearpage

\begin{figure*}[!t]
   \begin{tabular}{cc}
      \resizebox{7.cm}{!}{\includegraphics{f10}}
     &\resizebox{7.cm}{!}{\includegraphics{f11}}\\
   \end{tabular}
   \caption[]{One-dimensional FOCAS spectrum of the most luminous object in the field showing clearly the signature of the \ion{He}{2} line at 4686 \AA, which proves that it is the optical counterpart of the ULX. The other annotated emission lines are residuals of the non-perfect subtraction process of the inhomogeneous nebula.}
   \label{spectres_1D_contrepartie_hoixx1}
\end{figure*}

\clearpage

\clearpage

\clearpage

\clearpage

\clearpage

\clearpage

\clearpage

\clearpage

\clearpage

\clearpage

\begin{table*}
\begin{threeparttable}
\caption[]{Magnitudes of the ULX Counterpart and of its Close Object in the $V$ Filter Derived from SUBARU/FOCAS and {\it HST}/ACS Observations}
\label{tab_magnitudes_hoix_variabilite}
\small
\centering
\begin{tabular}{lcccc}
\hline\hline
Telescope  & Date 	       & $V_{\mathrm{ULX}}$ & $V_{\mathrm{close \; object}}$ & $V_{\mathrm{total}}$   \\
\hline
SUBARU/FOCAS    & 2003 Jan 25  & ...  & ...  & 22.62 $\pm$ 0.02 \\
SUBARU/FOCAS    & 2003 Jan 26  & 22.71 $\pm$ 0.037  &  24.81 $\pm$ 0.26  & 22.56 $\pm$ 0.26 \\
SUBARU/FOCAS    & 2003 Jan 26  & 22.68 $\pm$ 0.015 &  ... & ... \\
SUBARU/FOCAS    & 2003 Feb 9  & 22.67 $\pm$ 0.025  &  24.49 $\pm$ 0.12  & 22.48 $\pm$ 0.12 \\
\hline
\hline
{\it HST}/ACS   	& 2004 Feb 4  & 22.609 $\pm$ 0.024  &  24.54 $\pm$ 0.03  & 22.44 $\pm$ 0.04 \\
{\it HST}/ACS   	& 2004 Mar 25  & 22.745 $\pm$ 0.013  &  24.50 $\pm$ 0.02  & 22.55 $\pm$ 0.02 \\
\hline
\end{tabular}
\normalsize
\begin{tablenotes}
\item {\bf Note.} Magnitudes are expressed in the Johnson-Cousins system ($UBVI$).
\end{tablenotes}
\end{threeparttable}
\end{table*}

\clearpage

\acknowledgments

We thank Dr. Takeshi Go Tsuru for his help during the SUBARU/FOCAS observations. We thank the anonymous referees for their comments which improved this paper. F.G. thanks Jeanette Gladstone, Erik Hoversten, Roberto Soria, and Luca Zampieri for discussions.\\
This paper is based in part on data collected at SUBARU Telescope, which is operated by the National Astronomical Observatory of Japan and on observations made with the NASA/ESA {\it Hubble Space Telescope}, obtained from the data archive at the Space Telescope Institute. STScI is operated by the association of Universities for Research in Astronomy, Inc. under the NASA contract  NAS 5-26555.
This paper is based also on observations obtained at the Gemini Observatory (acquired through the Gemini Science Archive), which is operated by the Association of Universities for Research in Astronomy, Inc., under a cooperative agreement with the NSF on behalf of the Gemini partnership: the National Science Foundation (United States), the Science and Technology Facilities Council (United Kingdom), the National Research Council (Canada), CONICYT (Chile), the Australian Research Council (Australia), Ministério da Ciência e Tecnologia (Brazil) and Ministerio de Ciencia, Tecnología e Innovación Productiva (Argentina).
This research has made use of data obtained from the {\it Chandra} Data Archive, and software provided by the {\it Chandra} X-ray Center (CXC) in the application packages CIAO (v.4.1.2).

\bibliography{ms}
\bibliographystyle{apj}

\clearpage

\end{document}